\documentstyle[prb,preprint,aps,psfig]{revtex}    
\begin{document} 
\draft 
\newcommand{\no}{\nonumber} 
\newcommand{\beq}{\begin{equation}} 
\newcommand{\eeq}{\end{equation}} 
\newcommand{\beqa}{\begin{eqnarray}} 
\newcommand{\eeqa}{\end{eqnarray}}
 
\title{Self-consistent scattering description of transport \\ 
in normal-superconductor structures}

\author{J. S\'anchez-Ca\~nizares and F. Sols} 
\bigskip 
\address{ 
Departamento de F\'{\i}sica Te\'orica de la Materia Condensada, C-V, 
and\\ 
Instituto Universitario de Ciencia de Materiales ``Nicol\'as 
Cabrera''\\ 
Universidad Aut\'onoma de Madrid, E-28049 Madrid, Spain}

\maketitle 
\begin{abstract} 
We present a scattering description of transport in several 
normal-superconductor structures. We show that the related requirements of
self-consistency and current conservation introduce qualitative changes in the
transport behavior when the current in the superconductor is not
negligible. The energy thresholds for quasiparticle propagation in the
superconductor are sensitive to the existence of condensate flow ($v_s\neq 0$).
This dependence is responsible for a rich variety of transport regimes,
including  a voltage range in which only Andreev transmission is possible at
the interfaces, and a state of gapless superconductivity which may
survive up to high voltages if temperature is low. 
The two main effects of current 
conservation are a shift
towards lower voltages of the first peak in the differential 
conductance and an
enhancement of current caused by the
greater availability of charge transmitting scattering channels. 
\end{abstract} 
\vspace{.5cm} 
 
 
\vspace{.5cm} 
\narrowtext 
 
\section{Introduction} 
 
During the last few years, quantum transport in superconductor structures  
has been the object of renewed attention. 
Interest in this topic has grown as an extension 
of the research on electron transport in normal mesoscopic systems. 
A considerable amount of theoretical and experimental work has  
been devoted to coherent transport of quasiparticles in small 
structures containing normal and superconductor elements, \cite{hekk95} 
with the characteristic presence of phase coherent Andreev  
reflection.\cite{andr64,been95} 
Theoretical studies of superconducting  
transport in inhomogeneous structures based on a quasiparticle scattering  
picture are conventionally performed within the framework of the  
Bogoliubov - de Gennes (BdG) equations: \cite{dege66}
\beqa 
\left[ \begin{array}{cc} 
H_0 & \Delta \\ 
\Delta^* & -H_0^* \end{array} \right] 
\left[ \begin{array}{c} u_{n} \\ 
v_{n} \end{array} \right] 
= \varepsilon_{n} 
\left[ \begin{array}{c} u_{n} \\ 
v_{n} \end{array} \right], 
\eeqa 
where $[u_n,v_n]$ and $\varepsilon_n$ are the wave function 
and energy of quasiparticle $n$. In principle, the BdG equations 
must be solved in a 
self-consistent manner, so that the gap function $\Delta({\bf r})$ 
is required to satisfy the condition 
\beq 
\Delta = g \sum_{n} u_{n} v_{n}^*  
(1 - 2 f_{n}), 
\eeq 
$g$ being the electron-phonon coupling constant. 
The implementation of the self-consistency condition (2) has often  
been neglected in the literature on superconducting transport, 
and this omission has not always been entirely or explicitly justified. 
The role  of self-consistency in theoretical descriptions 
of superconducting transport has attracted some attention recently. 
Most importantly, it has been shown  that current conservation 
is generally guaranteed only within a self-consistent  scheme.
\cite{furu91,bagw94,sols94,ferr90} This result may be viewed 
as a particular case of a more general theorem in quantum statistical 
mechanics stating that conserving  
approximations must satisfy the mean-field equations.\cite{baym61} 
 
In a transport context, the requirement of current  conservation can be
satisfied only if the condensate carries  a finite amount of
current,\cite{sanc95} for which a nonzero phase gradient  is needed:  $\nabla
\varphi \neq 0$, with $\Delta  \equiv |\Delta| e^{i\varphi}$.  In the
asymptotic region,  the superfluid velocity $v_s \equiv \hbar \nabla
\varphi/2m$  acquires a uniform value, so that the gap behaves as \beq
\Delta({\bf r})=|\Delta|e^{2iqx}, \eeq where $q \equiv mv_s/ \hbar$ is half 
the Cooper pair momentum and $x$ is the longitudinal coordinate  in the lead.
The occurrence of a  nonzero  $q$ in the case  of equilibrium superconducting
flow is a basic result  that has been known for a long time,
\cite{dege66,bard62,tink96} and a related  study has been recently undertaken
by  Bagwell.\cite{bagw94}  In Refs. \cite{dege66,bagw94,bard62,tink96},  the
quasiparticles, whose dispersion relation is modified by a finite $q$, 
\cite{dege66} are assumed to be in equilibrium among  themselves and with
respect to the lattice at rest.  With this assumption, the  standard
Ginzburg-Landau (GL)  theory is derived from the   microscopic BdG equations
for  systems with a nonzero current  ${\bf j}=(e\hbar/m)  |\psi|^2 \nabla
\varphi$,  where $\psi$ is the (small) order parameter.  The inclusion of a
nonzero $q$ is essential to describe the crossover  from the Josephson   effect
between two weakly coupled superconductors  to bulk flow in a  single perfect
superconductor.\cite{sols94} The mismatch between  a moving condensate and a
population of   quasiparticles in equilibrium with the solid at rest is 
responsible for the fast suppression of superconductivity as the superfluid 
velocity reaches the depairing value $v_d\equiv \Delta_0/\hbar k_F$ 
($\Delta_0$ is the zero temperature, zero current gap and $k_F$ the  Fermi
wavevector). In a more general transport context, quasiparticles may  not be in
equilibrium with the condensate nor among themselves.  Following the work by
Blonder, Tinkham, and   Klapwijk (BTK), \cite{blon82,klap94} a number of
papers have dealt with scattering descriptions of   transport at finite bias in
hybrid normal-superconductor (NS) structures.  However, work on nonlinear  
transport through NS interfaces has generally not included  the effect of 
moving Cooper pairs. The purpose of this article is  to analyse 
{\it the combined  
effect of a nonzero superfluid velocity  
and a nonequilibrium  
population of quasiparticles}.  Traditional work on nonequilibrium
superconductivity \cite{gray81} has indeed contemplated the combination of  
finite   superfluid velocities and electric fields. However,  such studies have
been usually  limited to the GL regime and to semiclassical descriptions of
quasiparticle   transport. Here we attempt to present a  microscopic study
based on the self-consistent resolution of the BdG equations.  Because it
relies on  a quasiparticle scattering picture, this approach is similar in
spirit to that of recent works on mesoscopic
superconductivity,\cite{lamb91,been92,taka92} where the concepts of normal
mesoscopic transport \cite{land57,butt86} have been extended to include the
presence of superconducting elements.\cite{hekk95} 

A scattering description of transport in a realistic NS 
structure with an exact  
implementation of self-consistency is in general a demanding numerical  
task. In order both to lighten the required numerical work (and thus 
increase the scope of our study) and to isolate the  
essential physical facts, we have introduced several simplifying  
assumptions: (i) the gap amplitude $|\Delta|$ is assumed to be stepwise  
uniform; (ii) the self-consistency condition (2) is implemented   
asymptotically throughout the superconductor; and (iii) quasiparticle multiple  
scattering by more than one interface is assumed to be quasi-elastic and  
incoherent.  

In the present work, we present a detailed  analysis of the role   of current
conservation in nonlinear transport through hybrid NS  structures,   within the
framework of the model outlined above. In particular, we discuss   the nature
of the assumptions involved, describe the numerical method,   comment on some
conclusions that can be inferred from general   considerations, and present
specific results for the current-voltage   characteristics in several types of
NS structures.  Some preliminary results   for the NS interface and a
(symmetric) NSN structure were presented in Ref.   \cite{sanc95} The combined
effect of self-consistency and   nonequilibrium has also been studied by Martin
and Lambert   \cite{mart95} assuming coherent scattering and  implementing
condition (2) locally for NSN structures  in which one interface is perfectly
transmissive. Self-consistency at a NS interface in the case of zero current
was already studied by McMillan, \cite{mcmi68} who found that, at the scale of
the coherence length ($\xi_0\equiv \hbar v_F/\pi \Delta_0$), the gap shows a
depression on the S side and a finite discontinuity at the interface. 
More recently,
Bruder \cite{brud90} 
has performed a similar calculation for the case of anisotropic superconductors.
A systematic study of the self-consistent gap profile in a NSN structure for
different values of the current and the barrier strength will be presented
elsewhere.\cite{sanc97}

We wish to answer the questions: `What are the effects of 
self-consistency?' 
and `when are they important?'. While the bulk of this paper is devoted to 
the first question,  an answer to the second question can be presented in 
rather simple terms. A clear result that emerges from the work presented here, 
as well as from related papers,\cite{bagw94,sanc95,mart95} is 
that, at zero temperature, self-consistency is important when the superfluid  
velocity $v_s$ is comparable to the depairing velocity $v_d$.
Since $v_s$ must  
be obtained self-consistently after an iterative resolution of the BdG  
equations, a general relation that yields $v_s$ in terms of the applied 
voltage seems difficult to find. However, some conclusions can be reached 
from the following considerations: Let us  
consider a generic NSN structure, similar to any of those depicted 
in Figs. 1  
c-e (the argument could be easily generalized to structures 
with one superconducting terminal,  like those in Figs. 1 a,b), 
with the possible inclusion of some additional 
quasiparticle scattering within  
the N and S regions. When the applied voltage $V$ is sufficiently low, 
Andreev reflection  
is the only charge-transmitting channel. 
Incident electrons are reflected as holes, pumping 
Cooper pairs into the superconductor. In this low voltage regime, 
and at zero temperature, all the current 
in S is carried by the the condensate, with a value \cite{dege66} 
\beq 
I_{S} = ev_s n A, 
\eeq 
$A$ being the transverse area and $n$ the electron density. 
On the other hand, the current in the normal leads can be written as 
\beq 
I_{N} = \frac{2e^2}{h} N_{\rm ch}T_0 V, 
\eeq  
where $N_{\rm ch}$ is the number of available channels 
at the Fermi level, and  
$T_0 \le 1$ is the average transmission probability for 
electrons coming from the normal leads. Current  
conservation requires  $I_S = I_N$. As the voltage increases, 
the condensate  
carries more current and self-consistency effects become 
important when 
$v_s \alt v_d$. A simple analysis reveals that this condition 
translates into 
\beq 
T_0 V \alt \alpha \Delta_0/e, 
\eeq 
where $\alpha$ is a number of order unity: $\alpha=2$, 
$\pi/2$, and 4/3 in one,  
two, and three dimensions, respectively. Since the most 
interesting physics occurs for $V$  comparable to $\Delta_0/e$ 
(at higher voltages, the  
superconducting features tend to become marginal), 
one infers from (6) 
that the effects of finite superfluid flow can only be important 
for $T_0$ close to unity. Thus, we may formulate the following criterion:
{\it At voltages of  
order $\Delta_0/e$, the effects of a finite condensate 
flow are important in  NSN
structures of uniform width containing transmissive NS interfaces}.
The condition of equal width guarantees  
the possibility of achieving high currents in the superconductor at the  
voltages of interest. The argument  
above can be generalized to the case where the N and S 
wires have different widths. Let $W_S$ and $W_N$ be the  
widths of the superconductor and of the {\it narrowest} normal wire.  
One can show that, 
(i) when the superconductor is narrower than both normal wires  
($W_S<W_N$), the condition (6) applies identically with $T_0$ 
defined differently but 
also bounded by a maximum value of one, reserved for 
structures with ideal contacts, and (ii), when S is wider than at least one  
normal wire ($W_S>W_N$), the r.h.s. of (6) must be multiplied by 
a factor  
$W_S/W_N>1$, thereby raising the scale of voltages at which $v_s$ 
becomes comparable to $v_d$ (dilution effect) and hence making
self-consistency unimportant.
 
It is interesting to note that the criteria formulated above  exclude those
structures   with poor transmission from one of the normal sides as possible 
candidates to   display the current-conservation effects discussed in this
article.  In particular,   we cannot expect the physics of finite flow to be
relevant  in structures where   electrons and holes experience weak
localization in the N region,  since that   requires a small average
transmission: $T_0 \simeq \pi l/2L \ll 1$,  with  $l$   the impurity mean free
path and $L$ the length of the disordered normal   region.\cite{been91} This
result holds even at the resistance minima   characteristic of coherent Andreev
reflection, because these are still comparable   to the classical normal state
resistance.\cite{mels94} 

When studying nonlinear transport in superconducting structures  
within a current-conserving description, some ideas developed in  
conventional nonequilibrium superconductivity \cite{gray81} are  
necessarily revisited, albeit from a new perspective based on a  
scattering picture. For instance, we reencounter the phenomenon of charge  
imbalance, whereby quasiparticles are not in equilibrium with the  
condensate.\cite{clar72} In particular we find that lack of complete  
thermalization strongly reduces the efficiency with which 
the presence of quasiparticles tends to destroy 
superconductivity, and from Eq. (2) we identify a general 
reason for this behavior. 
The fact that a nonequilibrium  
population of quasiparticles can 
strengthen superconductivity has 
been shown for SNS structures both experimentally 
\cite{blam90} and theoretically.\cite{klap93} 
In Refs. \cite{blam90,klap93} it is noticed 
that a nonzero superconducting order parameter 
may exist in the N segment at $T>T_c$ 
provided that equilibration of quasiparticles is 
incomplete. Here we predict a similar 
phenomenon by which, due to quasiparticle 
nonequilibrium, a superconductor 
(with $T<T_c$) may have 
$v_s>v_d$ and still display a nonzero order parameter.

The existence of a nonzero flow in the condensate gives rise to a rich  
variety of transport regimes. At low  
enough voltages, current is solely transmitted by Andreev reflection. 
As the voltage increases, so does $v_s$, and the quasiparticle  
dispersion relation becomes correspondingly distorted (compare the 
insets of Figs.  
2a and 2b). This causes a {\it lowering of the threshold voltage for 
quasiparticle transmission}, which occurs at $V= \Delta_{-}/e$, 
where $\Delta_{\pm}\equiv  \Delta_0 \pm \hbar q  
v_F$. Above this voltage, Andreev transmission (AT) converting 
electrons into quasiholes becomes possible, while normal transmission 
(NT) as quasielectrons is still not available if $V<\Delta_+/e$. 
Some properties of this regime in which AT is the only  
quasiparticle transmission mechanism have been analysed in  
Ref. \cite{sanc96}, the most important being the insensitivity of
transport to the presence of impurities, since these can only
induce Andreev reflection and thus cannot degrade the current. 
As the voltage increases further, transport can evolve  
in different ways. For very transmissive structures, and at low  
temperatures, $\Delta_-$ may become negative and a state of gapless  
superconductivity (GS) is achieved (see the inset of Fig. 2c).
In less transmissive systems, $v_s$ remains low and some regimes
may be unstable. 

This paper is arranged as follows: In section II, we present  the model  
employed in our calculations, and discuss its range of  physical   validity.
In section III, we describe the various scattering  mechanisms which, with
different energy thresholds, can operate in hybrid NS structures. Section IV
deals with  some transport properties that can   be inferred from very general
considerations.  In sections V-IX,   we present and discuss specific numerical
results for several prototypical structures. Section V addresses
self-consistent   transport in the most basic structure: the NS interface. 
Several remarkable   results (such as a negative peak in the differential
conductance) are argued   to be not observable in practice. However, their
study paves the way for a deeper   understanding of transport in more realistic
structures  such as NSS'   (section VI), where S' is a wide superconductor with
$v_s$ always negligible, or a symmetric NSN (section VII).  The effect of
several transverse channels (still within the regime of
quasi--one-dimensional superconductivity) 
is studied  in section VIII,  where the interesting
conclusion is reached that the basic physics learned from one-dimensional
models remains valid. Section IX  is devoted to the asymmetric NSN structure,
and a final summary  is given  in Section X. This paper is complemented by
Appendix A,  where we show that there is a well-defined   asymptotic region in
the S wire where the gap amplitude,  as determined by the self-consistency
condition, becomes   independent of the the longitudinal coordinate.  This
result underlines the physical relevance  of results obtained within the
approximation  of asymptotic self-consistency.

\section{The model} 
 
Our goal is to solve the 
BdG equations for several structures of interest.
In Eq. (1), 
$H_0$ contains the kinetic energy plus a barrier 
potential of the form $V(x)=H\delta(x)$ at each NS 
interface. The parameter 
$Z=mH/\hbar ^2 k_F$ is a dimensionless measure 
of the barrier strength.\cite{blon82}
As in other scattering studies of transport, we choose a basis of 
retarded scattering states. Each quasiparticle $n$ 
is characterised by its energy $E$
and by a discrete index $\alpha$ labeling the {\it incoming} 
scattering channel.
Thus, quasiparticle $n\equiv (E,\alpha)$ is populated 
with probability $f_n=f_{\alpha}(E)= 
f_0(E- \tau \mu _{\alpha})$,  where $f_0(E)$ is the 
Fermi-Dirac distribution,
$\mu_{\alpha}$ is the chemical potential at the emitting 
reservoir, and 
$\tau_{\alpha}=1 (-1)$ for incoming electron (hole) channels 
(charge $e>0$ is assumed for electrons and a convention of 
$E>0$ for all $n$ is adopted). For superconducting terminals, 
the noninteger value of  the quasiparticle charge is not a 
problem because for them $\mu_{\alpha} \equiv 0$.
 
The total current can be written in terms of a sum 
over quasiparticle states: 
\beq 
I(x)=\frac{2e\hbar}{m} \sum_{n} \mbox{Im} \left[ 
f_{n} u_{n}^*(x)\nabla u_{n}(x) - 
(1-f_{n})v_{n}^*(x)\nabla v_{n}(x) \right], 
\eeq
where the factor $2$  accounts for spin degeneracy.
The condensate current is identified with the term which is 
independent of the occupation probabilities $\{f_n\}$:
\beq 
I_{\rm c} (x)=-\frac{2e\hbar}{m} \sum_{n} \mbox{Im} \left[ 
v_{n}^*(x)\nabla v_{n}(x) \right], 
\eeq
On the other hand, the quasiparticle component of the electric 
current is given by the term linear in $\{f_n\}$:
\beq 
I_{\rm qp}(x)=\frac{2e\hbar}{m} \sum_{n}  
f_{n} \mbox{Im} 
\left[ 
u_{n}^*(x)\nabla u_{n}(x) + 
v_{n}^*(x)\nabla v_{n}(x) \right] . 
\eeq 

In Appendix A, we show that,
except for unphysical oscillations at the 
scale of the Fermi wavelength,
the amplitude of the r.h.s. in Eq. (2) becomes 
independent of $x$  at distances from the scattering 
region much greater than $\xi_0$. Hence, $\Delta({\bf r})$
behaves asymptotically as in Eq. (3). This permits to describe transport
in terms of a well defined scattering problem, since 
the scattering channels are known exactly. They are given 
by the solutions of the BdG equations for a perfect superconductor 
with a gap of the form (3).\cite{dege66,tink96} 
The quasiparticle dispersion relation is \cite{dege66}
\beq 
\varepsilon(k) =\frac{\hbar^2kq}{m} \pm \left[\left(
\frac{\hbar^2}{2m}\right)^2 
(k^2+q^2-k_F^2)^2 + |\Delta|^2\right]^{1/2},
\eeq
and the coherence factors are
\beqa 
u(k) 
&=& 
\frac{|\Delta|} 
{[(\varepsilon_k-\xi_{k+q})^2+|\Delta|^2]^{1/2}}, \nonumber \\ 
v(k) 
&=& 
\frac{\varepsilon_k-\xi_{k+q}}{[(\varepsilon_k-\xi_{k+q})^2 
+|\Delta|^2]^{1/2}}, 
\eeqa 
where $\xi_k=\hbar^2k^2/2m - E_F$. 

In terms of the scattering channels, Eqs. (2), (8), and (9) are rewritten as
\beq 
|\Delta| = \frac{g}{h} \int_0^{\hbar \omega_D} dE 
\sum_{\lambda}\frac{1}{\nu _{\lambda}} 
v_{\lambda} 
u_{\lambda} 
(1 - 2 \tilde{f}_{\lambda}), 
\eeq

\beq 
I_{\rm c}=\frac{e}{\pi m} \int_0^{\infty} dE 
\sum_{\lambda}\frac{1}{\nu _{\lambda}} 
(q-k_{\lambda}) 
v_{\lambda}^2, 
\eeq

\beq 
I_{\rm qp}=\frac{e}{\pi m} \int_0^{\infty} dE 
\sum_{\lambda}\frac{1}{\nu _{\lambda}} 
\tilde{f}_{\lambda} 
[ k_{\lambda} + q(u_{\lambda}^2 
- v_{\lambda}^2) ]. 
\eeq 
In Eqs. (12-14), the subindex $\lambda$ runs over all
channels available at energy $E$. These are
determined by the real solutions 
$k_{\lambda}$ to the equation $\varepsilon(k)=E$. 
$\nu _{\lambda}$ is the absolute value of the group velocity 
for channel $\lambda$, i.e., $\nu _{\lambda}=(1/\hbar)
|d\varepsilon(k_{\lambda})/dk_{\lambda}|$.
$u_{\lambda}$ is an abbreviation for $u(k_{\lambda})$,
and similarly for $v_{\lambda}$.
In the gap equation,
a standard cutoff at the Debye energy has been 
introduced to ensure a finite $|\Delta|$.
The occupation probabilities $\{\tilde{f}_{\lambda}\}$ are taken
$\tilde{f}_{\alpha}=f_{\alpha}$ for the incoming states $\alpha$, and
\beq 
\tilde{f}_{\beta}=\sum_{\alpha}f_{\alpha}|S_{\beta \alpha}|^2, 
\eeq
for the outgoing channels $\beta$, where $|S_{\beta\alpha}|^2$ 
is the  probability for a quasiparticle to be transmitted 
from $\alpha$ to $\beta$.

The term in the r.h.s. of Eq. (14) which is explicitly linear in $q$ may be
written in the form $(2e\hbar/m) q \delta Q^*$, where 
\beq 
\delta Q^* \equiv
\frac{1}{\hbar}\int _0 ^{\infty} dE \sum _{\lambda}\frac{1}{\nu _{\lambda}}
\tilde{f}_{\lambda} (u_{\lambda}^2 
- v_{\lambda}^2)
\eeq 
is referred to as {\it quasiparticle charge
imbalance}.\cite{blon82,peth79} In all the cases that we have considered, its
contribution to the quasiparticle current is  negligible, namely,
$(2e\hbar/m)q\delta Q^*/I_{\rm qp} < 10^{-4}$.

The transmission coefficients $S_{\beta\alpha}$ are 
obtained from the solution of the scattering problem. 
The effect of the delta barrier is expressed
through the matching conditions
\beqa 
\psi_L (0) &=& \psi_R (0), \nonumber\\ 
\psi_R'(0)-\psi_L'(0) &=& \frac{2mH}{\hbar^2}\psi(0). 
\eeqa
At this point we introduce the simplifying approximation that 
the gap amplitude is stepwise uniform, i.e., 
$|\Delta(x)|=|\Delta| \Theta(x)$. In this way, the calculation reduces 
to a generalization of the work by BTK  \cite{blon82} to include a 
nonzero value of the superfluid velocity which will be determined by the 
requirement of current conservation. 

To compute the transport properties of structures with  more than one
interface, such as NSN or NSS', we introduce  the assumption of {\it
incoherent} multiple scattering of  quasiparticles by the interfaces. We wish
to remark that  this hypothesis is equivalent to that which in normal
transport  leads to a picture where the contribution $(h/2e^2)R/(1-R)$  of a
given obstacle to the resistance \cite{land57} ($R$ is the reflection
probability) can be treated as an additive quantity.  In the case of
superconducting transport, the assumption of  incoherent scattering does not
lead to a simple additive rule  for the resistance because of the presence of
the condensate,  which introduces qualitative differences in the required
treatment. Within a picture of incoherent scattering, where probabilities 
--not amplitudes-- are added, the description of transport in a finite S
segment  does not differ much from the case where S is a semiinfinite lead. 
Eqs. (12-14) still apply if, for $\tilde{f}_{\lambda}$, we use \beq 
\tilde{f}_{\lambda}=\sum_{\alpha}f_{\alpha}T_{\lambda \alpha},  \eeq where
$T_{\lambda \alpha}$ is a characteristic concept of  Bayesian statistics:  it
is the probability that a quasiparticle found traveling in channel $\lambda$
entered the structure initially from  the incoming channel $\alpha$. 

As an example of how to compound scattering probabilities, 
let us consider the NSN structure. At the left interface, we may write
\beqa
\hat {P}(q)
\left[ \begin{array}{c} |e^+_{L}|^2 \\ 
|h^+_{L}|^2 \\ |e^-_S|^2 \\ |h^-_S|^2 
\end{array} \right]
\equiv 
\left[ \begin{array}{cc}
R_{ee} \ R_{eh} \ T'_{ee} \ T'_{eh} \\ 
R_{he} \ R_{hh} \ T'_{he} \ T'_{hh} \\ 
T_{ee} \ T_{eh} \ R'_{ee} \ R'_{eh} \\ 
T_{he} \ T_{hh} \ R'_{he} \ R'_{hh} 
\end{array} \right] 
\left[ \begin{array}{c} |e^+_{L}|^2 \\ 
|h^+_{L}|^2 \\ |e^-_S|^2 \\ |h^-_S|^2 
\end{array} \right] 
= 
\left[ \begin{array}{c} |e^-_{L}|^2 \\ 
|h^-_{L}|^2 \\ |e^+_S|^2 \\ |h^+_S|^2 
\end{array} \right], 
\eeqa 
where the matrix elements are probabilities 
corresponding to 
a NS junction with a phase gradient $2q$.
$e$ and $h$ refer to the type of quasiparticle
and $\pm$ indicates the sign of the group velocity,  positive if
motion is to the right. The array on the l.h.s. (r.h.s.) of Eq. (19) contains
the square moduli of the amplitudes for the incoming (outgoing) channels.
The prime indicates scattering for particles 
coming from the S side. At the right
interface, we have, analogously,
\beqa
\hat{P}(-q) 
\left[ \begin{array}{c} |e^-_{R}|^2 \\ 
|h^-_{R}|^2 \\ |e^+_S|^2 \\ |h^+_S|^2 
\end{array} \right]
= 
\left[ \begin{array}{c} |e^+_{R}|^2 \\ 
|h^+_{R}|^2 \\ |e^-_S|^2 \\ |h^-_S|^2 
\end{array} \right]. 
\eeqa
Combining Eqs. (19) and (20), one may derive a linear relation giving the 
unknown  probabilities
($|e^-_L|^2,|h^-_L|^2,|e^{\pm}_S|^2, |h^{\pm}_S|^2,
|e^+_R|^2,|h^+_R|^2$) in terms of the fluxes at the incoming
channels, namely, $|e^+_L|^2,|h^+_L|^2,|e^-_R|^2,|h^-_R|^2$. 
The quantities $|S_{\beta\alpha}|^2$ or
$T_{\beta\alpha}$ are obtained by assuming only one incoming channel
is populated. Then, by introducing (19) and (20) in Eqs. (12-14), 
the current and the gap can be computed.

Self-consistency at the NS interface is  implemented as follows:
At a given voltage, we solve for the scattering problem with guess values for  
$q$ and $|\Delta|$ and compute the current in the N lead from the 
trivial normal counterpart of Eq. (14). The gap and the current 
in the superconductor are obtained from Eqs. (12-14). 
The new value of $|\Delta|$ is used as an input for the next iteration. 
The new $q$ value is 
adjusted by requiring current conservation, 
$I_N=I_S=I_{qp}+I_c$, and exploiting the fact
$I_c$ has an explicit linear dependence on $q$.
With the new $q$ and $|\Delta|$ the scattering calculation
is performed again. The iterative process continues until
self-consistency is achieved. Then, for a given value of the applied voltage,
we have a pair of
values $(q,|\Delta|)$ yielding scattering results that, when
introduced in Eq. (2), give the same value of $|\Delta|$
and satisfy current conservation.
By varying the voltage, we
reproduce the self-consistent I--V characteristic of 
the NS structure. Generalization to structures with more than one interface
is made following the indications given in the previous paragraph.
The best guess values at a given step in
the iteration are those of the 
last I--V point calculated. 
Self-consistency is especially difficult to attain at 
the beginning of every new transport regime. 
It is easily lost when the guess values lie 
very far from the true ones.  For this reason, small 
voltage incremental steps (compared with the typical energy 
$\Delta_0$) must be introduced to ensure success 
in the next I--V point.

The assumption of incoherent elastic scattering requires the existence of some
quasi-elastic dephasing mechanism  that randomizes the phase with negligible
energy degradation. Thus, if  $l_{\phi}$ is the  dephasing length and $L$ is
the length of the superconductor,  we require $L \gg l_{\phi}$. On the other
hand, $L \gg \xi_0$  is needed for $|\Delta|$ to reach its asymptotic value. We
also assume that quasiparticles do not relax within the superconductor among
themselves nor with the superconductor, except for the indirect coupling
required by the self-consistency condition. Our calculations show this lack of
equilibration  is responsible for notable transport properties. Relaxation is
expected to be negligible when the charge imbalance \cite{gray81,clar72}
relaxation time,  $\tau_{\varepsilon}$, is much longer than the average
residence time $\tau_r$. In summary, we find that  $L \agt  \mbox{max} \{\xi_0,
l_{\phi}\}$ and $\tau_{\varepsilon}  \gg \tau_r$ must be satisfied for our
model to be meaningful. For not very reflecting interfaces, $\tau_r  \simeq
L/v_F$, so that simultaneous fulfilment of  the two above conditions requires
$\tau_{\varepsilon}  \gg \hbar /\Delta_0$. This strong inequality is realizable
at sufficient  low temperatures.\cite{likh79} Another implicit assumption is
that the condensate has a uniform chemical potential. This is necessarily the
case when phase-slips are effectively forbidden. For this reason we need
operation far from the critical temperature ($T\ll T_c$).\cite{lang67} 

\section{Scattering mechanisms} 
    
Transport properties of a hybrid normal-superconductor structure are  strongly
dependent on the microscopic scattering of quasiparticles. At an NS interface
there are  four characteristic scattering processes. \cite{blon82} An electron
coming from the N lead may undergo normal  reflection (NR), normal transmission
(NT) as a quasielectron,  Andreev reflection (AR) as a hole, and Andreev
transmission (AT) as a quasihole.  Quasiparticle transmission into the
superconductor can only  occur above certain energy thresholds.  The AT
process takes place for $E>\Delta_-$ while NT requires $E>\Delta_+$. For
sufficiently large $q$, we have seen that $\Delta_-$ may become negative before
the $\Delta_+$ threshold is reached. 
In such a case, NT into a nonconventional branch may
occur for energies $0<E<|\Delta_-|$ (see inset of Fig. 2c). If $(Q_1,Q_2)$ are
the charges transmitted to the quasiparticle ($Q_1$) and condensate  ($Q_2$)
current components of the superconductor  in a scattering event for an electron
of charge $e$ coming from the  N lead, we have $(0,0)$ for NR, $(e,0)$ for NT,
$(0,2e)$ for AR,  and $(-e,2e)$ for AT. The charge going into the condensate is
adjusted to preserve current  conservation, taking into account the
quasiparticle charge that is  reflected into the N lead. The charge $2e$
absorbed by the condensate  in the two Andreev processes can only be carried
away from the NS  interface if the Cooper pairs move, i.e., if $q \neq 0$.
Since AR  and AT contribute to loading the condensate, higher values of $q$ 
are required as more of these two events occur, if current is to be  conserved.
Because Andreev events are the only possible processes  at low energies, the
superfluid velocity is expected to increase  strongly with voltage in the low
voltage region. 

The parameter $Z$ is a dimensionless measure of the barrier scattering 
strength. If the S lead is made normal ($\Delta_0=0$), then only 
NT and NR occur, with probabilities
\beq 
|t|^2=1-|r|^2 = \frac{1}{1+Z^2}, 
\eeq
where $t$ ($r$) is the transmission (reflection) amplitude. 
The amplitudes for the scattering processes at an NS 
interface can be written exactly in terms of the one-electron 
coefficients $r$ and $t$, provided that $q=0$ and the standard 
approximation is introduced that, for matching purposes, the
quasiparticles wave vectors are replaced by $\pm k_F$. 
If $a,b,c,d$ are the amplitudes for AR, NR, NT, and AT, respectively, 
one obtains
\beqa 
a &=& u_0v_0 |t|^2/D, \nonumber\\ 
b &=& (u_0^2-v_0^2)r/D, \nonumber\\
c &=& u_0 t/D, \nonumber\\
d &=& v_0 r^*t/D, 
\eeqa
where $D=u_0^2-v_0^2|r|^2$, and $u_0^2=1-v_0^2=
[1+(E^2-|\Delta|^2)^{1/2}/E]/2$. For $E<|\Delta|$ the square 
root is replaced by $i (|\Delta|^2-E^2)^{1/2}$. Eq. (22) 
applies to asymmetric barriers provided that $r$, $t$ 
denote amplitudes for electrons coming from the left. 
\cite{commscatt} Explicit reference to amplitudes from 
the right has been removed by invoking unitarity of the scattering matrix. 
Eq. (22) shows in a transparent manner that AR involves 
the transmission of two electrons, with the determinant 
$D$ accounting for extra multiple reflections at the interface. 
AT is an involved process which requires transmission of the electron 
with a simultaneous internal reflection of a hole 
within the superconductor. 
When current conservation is neglected ($q=0$), 
AT is not important. 
Being available only for $E>\Delta_0$, it is always 
overshadowed by the more probable NT process (note that $|d| < |c|$),
which has the same energy threshold.\cite{blon82} 
The situation changes drastically when $q$ becomes nonzero, 
because AT and NT open at different energies. 
In particular, we shall see that {\it there 
exists a range of voltages in which AT is the only 
quasiparticle transmission mechanism}.
Note, however, that AT is forbidden in very 
transmissive interfaces with
$|r| \simeq 0$.

In Fig. 2, we plot the four scattering probabilities for a barrier of
intermediate strength ($Z=1$) and several values of the superfluid velocity.
The insets show the schematic dispersion relation in each case. For greater
accuracy, the BdG equations have been solved exactly. For completeness, we show
in  Fig. 2a the scattering curves for $v_s=0$,  retrieving results implicitly
given in Ref. \cite{blon82}. Fig. 2b deals with a moderate  superfluid velocity
($v_s=0.7v_d$)  for which the initial threshold at  $E=\Delta_0$ splits into
$\Delta_{\pm}$.  In the range $\Delta_- \leq E \leq \Delta_+$, we observe that
AT occurs, while NT is not possible until $E \geq  \Delta_+$. In this energy
range AT is therefore the only mechanism of quasiparticle transmission. In Fig.
2c, we present results for the case $v_s \geq v_d$. Formally, a negative
$\Delta_-$ signals the onset of gapless superconductivity. With proper
inclusion of self-consistency, it has been shown that {\it equilibrium} GS
cannot exist within a perfect one-dimensional wire,  \cite{bagw94} and only in
a very small range of $v_s$ values if  three-dimensionality is taken into
account.\cite{bard62}.  Lack of equilibration changes the picture
qualitatively,  permitting GS to be a well-defined, stable transport regime.
\cite{sanc95} We will  return to these scattering mechanisms, as well
as to the transport regimes  they define, when discussing numerical results for
transport in specific structures.

\section{General remarks} 
 
In this section we pay attention to some general features  of the model that
will be helpful in the discussion  of results later in this paper. If we focus
on the condition of asymptotic self-consistency (see Eq. 12), we note that
the coherence factors $u_{\lambda},v_{\lambda}$, defined in Eq. (11), are both
real because $k_{\lambda}$ is real. By factoring out the $x$ dependence 
(which goes like
$e^{2iqx}$) in (3), we obtain an equation for $|\Delta|$.
Clearly, the properties  of $u(k)$ and $v(k)$ will affect the
magnitude of the gap.

Note the double sign appearing  in Eq. (10) for the dispersion relation. When
the superfluid  velocity is moderate ($v_s<v_d$), only the plus  sign leads to
a positive energy, and this is true
for any $k$-vector that may be considered. \cite{commconv}
However, when the superfluid velocity exceeds $v_d$, both signs lead  to a
positive quasiparticle energy in the neighborhood of  $+k_F$, while none of
them can yield a positive energy in the vicinity of $-k_F$. This is the GS
regime shown in the inset of Fig. 2c. A simple analysis reveals
that $v(k)$  is positive for the plus solution of Eq. (10), and negative for 
the minus solution. Since $u(k)$ is always positive, we conclude that 
$\mbox{sgn} [u(k)v(k)]<0$ for states in the new, unconventional branch.  This
implies that the presence of these new quasiparticles tends to  cancel the
contribution of the existing conventional quasiparticles to the r.h.s. of the
gap equation (12),  i.e., with their presence they tend to {\it reinforce}  the
gap.\cite{sanc95,commgap} This behavior contrasts  markedly with that found for
thermal quasiparticles.  In such a case, the states of $E\simeq 0$ becoming
available  with the onset of GS are populated with probability  $f_{\lambda}
\simeq \frac{1}{2}$, which contributes  to depress the gap. 
\cite{tink96}  An important physical consequence is that, in the absence 
of quasiparticle equilibration, {\it a nonzero $|\Delta|$ can survive up to  
high voltages} (we have explicitly checked $eV \agt 5k_BT_c$  in our
calculations). This effect manifests itself as inefficient heating  in the
presence of high voltage transport. Lack of heating has been observed,
\cite{rodr94,poza95} but it could also be due to a mere
dilution effect. The GS state disappears quickly with
increasing temperature.  We shall see that, even at zero temperature, the
directional  randomization caused by the confining effect of two opaque 
barriers can also contribute to depress the gap.  For not very long
superconductors ($L \sim \xi_0$),  the penetration of incoming quasiparticles
below the gap  can also be a cause of gap reduction. \cite{mart95,sanc97}. 

We end this section with some thermodynamic considerations.  Although our
system is not in equilibrium, we take advantage  of the fact that the state of
the system can be characterized  by the properties of the emitting reservoirs,
each of which is  in internal equilibrium. The population of quasiparticle
states and the properties of the condensate are determined by  the chemical
potentials at the reservoirs as well as  by the current-conserving  scattering
processes within the structure. This picture of local equilibrium permits  us
to adopt a statistical definition for the free energy:
\beq
{\cal F} = \frac{1}{h} \int_0^{\hbar \omega _D} dE
\sum_{\lambda}\frac{1}{\nu_{\lambda}}\{E_{\lambda}(\tilde{f}
_{\lambda}-v_{\lambda}^2)+k_B T [ \tilde{f}_{\lambda}
\log \tilde{f}_{\lambda} + (1-\tilde{f}_{\lambda})
\log (1-\tilde{f}_{\lambda}) ] \}.
\eeq
This expression agrees with Eqs. (23) and (24) 
of Ref. \cite{bagw94} for a uniform superconductor
with equilibrium flow, if the Fermi-Dirac distribution is replaced
by $\tilde{f}_{\lambda}$ as defined in Eq. (15). In our calculations, 
we have checked that the free energy so defined is smaller 
than that which would be obtained by setting 
$\Delta$ to zero ($ {\cal F}_{\rm S} <{\cal F}_{\rm N})$.

\section{I--V at NS} 
 
In Figs. 3 and 4, we present numerical results  
for several NS junctions of  different barrier strengths.
Here, as in the calculations shown in the following three sections, we
have taken a bandwidth 
of $E_F\!=\!5$ eV, a cutoff energy of 
$\hbar \omega_D \!=\! 0.1$ eV, 
and a zero temperature, zero current gap of 
$\Delta_0\!=\!1$ meV. 
These numbers yield a critical temperature $T_c\!=\!6.6$ K. 
Unless otherwise stated, the results shown throughout
this paper correspond to zero temperature.
In Fig. 3a and 3b, we plot the total current  and its quasiparticle 
component versus applied voltage.
Non self-consistent results are shown for comparison.
Fig. 4 shows the superfluid velocity and the gap amplitude
determined self-consistently for each value of the voltage.
In the fully transparent case ($Z=0$), the current $I$ increases linearly
as $(4e^2/h)V$ for 
$V<\Delta_0/e$. In this low voltage regime, current
is transmitted at the interface by AR and it is carried in S by the
condensate. In this case, the AT channel 
is forbidden because $r=0$ for $Z=0$. 
Fig. 4a shows that the
superfluid velocity also rises linearly until it 
reaches the depairing value $v_d$, point at which the system enters 
the GS regime. This transition is marked by a sharp
drop of the current to values close to $(2e^2/h)V$ corresponding to those
of an `ideal' NN interface, i.e., a perfect N wire.
In spite of 
the enhancement of $v_s$, the reduction in 
$|\Delta|$ is so strong that the condensate current becomes negligible.
$|\Delta|$ decreases because, with GS,
a new hole-like branch appears that is however left empty because 
no quasiparticles come from the S side. On the other hand, 
the emerging electron-like branch is strongly filled (NT has 
a probability close to one) but that simply replaces the effect 
of the empty conventional hole-like branch that has sunk below $E=0$.
Because the vanished empty conventional electron-like branch 
has been replaced by an empty unconventional hole-like branch, 
the net effect is a stronger cancellation on the r.h.s. 
of Eq. (12) and a corresponding reduction of the gap.
$|\Delta|$ is also diminished because, being overshadowed by NT, 
AR is no longer the dominant charge transmitting mechanism, and less 
current has to be carried by the condensate. We do not have a 
simple physical explanation to the small increase of 
$v_s$ at the onset of GS. It is noteworthy that, in these conditions, 
the system still prefers energetically to have a small finite 
value of $|\Delta|$.

Except for the existence of AT, the situation with $Z=0.5$ is very similar to
the transparent case. At $v \equiv eV/\Delta_0 \simeq 0.6$, a slight increase
in the differential conductance ($dI/dV$) is caused by the opening of the AT
channel. A characteristic feature of AT is that the quasiparticle current is
negative (see Fig. 3b), since it is transmitted by holes. The total current
increases, however, because a new Cooper pair is added  to the condensate in
each scattering event. This translates into a faster  increase (as a function
of voltage) of the superfluid velocity (see Fig. 4a).  Fig. 4b shows a slight
decrease of the gap caused by the opening and (only) partial filling of a
conventional channel. A much more drastic reduction  of the gap takes place
with the onset of GS at $v \simeq 1.2$, point at  which the quasiparticle
current begins to increase with voltage.  The gap decay is even stronger  than
in the transparent case because the emerging electron-like  branch is now less
populated, since more electrons are normally reflected.

For $Z=1$ and $Z=2$, the barrier is less transmissive and the total current
becomes smaller (note that in all the figures 
the current is plotted in units of
$I_0\equiv 2e\Delta_0|t|^2/h$). Because of this, the superfluid  velocity is
also smaller, and the branch distortion  is less important. The AT channel is
therefore  opened at higher voltages ($v \simeq 0.8$ for $Z=1$ and  $v \simeq
0.95$ for $Z=2$). Still, there are some similarities with the case of $Z=0.5$,
namely, an  increase of $v_s$ and of the condensate current, a negative 
quasiparticle contribution to the current, and  a small $|\Delta|$ decay due to
the low population of conventional quasiparticles  caused by poor
transmission.\cite{sanc95} As the voltage increases in the AT regime, GS is not
eventually achieved because, being $v_s$ low, the threshold $\Delta_+$ for
conventional NT is reached before $\Delta_-$ becomes negative. Unlike in the
GS case, the onset of this new  regime is not accompanied by a drop in the
total current. Only a slight change in the slope of the current (visible at
$Z=1$) reveals the existence of NT. A clearer signature appears in the plot of
$I_{\rm qp}$, where the slope changes from negative to positive (at $v \simeq
1.6$ for $Z=1$ and $v \simeq 1.1$  for $Z=2$). Conventional NT is accompanied
by very moderate  reductions of $v_s$, $I_{\rm c}$, and $|\Delta|$, as well as
by a rise in  $I_{\rm qp}$. The weak  dependence of $v_s$ and $|\Delta|$ on
voltage in the high voltage region is due to the poor  transmittivity of the
interfaces.

Fig. 3a predicts a sharp decrease in the current through transmissive
structures at a critical voltage value, 
and one wonders why it has never been observed. 
Typical measurements of the excess current predicted by the 
BTK model (dotted lines) have been performed in structures with a wide 
superconducting electrode (S'), with  
or without a 
narrower superconducting segment (S) connecting 
the N and S' leads. A wide superconducting lead S' may be 
described theoretically by a pair momentum that remains close 
to zero ($q\simeq 0$) because of current dilution.
The NS' interface is well described by the BTK 
model without requiring self-consistency. In the next section 
we show that  the NSS' structure does also display an excess 
current that is even greater than in the NS' case.

\section{I--V at NSS'} 
 
In Fig. 5, we show the I--V characteristic for a NSS' structure. In S, the
values of $q$ and $|\Delta|$ are adjusted self-consistently,  while in S',
$q=0$ and  $|\Delta|=\Delta_0$ are always taken. We assume a clean contact at
the SS' interface ($Z_S=0$) with a zero voltage drop. Non self-consistent (NS')
results are also shown for comparison. We obtain the desired excess current, 
thus recovering qualitative agreement with experiments.  Let us discuss first 
the $Z=0$ case. As in the NS interface, GS is reached for $v=1$ and a  large
current flux is carried by the unconventional quasielectron branch. 
However, these quasielectrons {\it can only be Andreev reflected at the SS'
interface}, and, after traveling through S, the quasiholes are normally 
transmitted at the NS interface with probability close to one.
Thus, {\it Andreev reflection is enhanced} by the presence of S. The
result is that, being AR still very probable, the slope of $I$
continues to  be $4e^2/h$ for $v>1$. The difference in behavior with the NS' 
(dotted lines) and NS (Fig. 3a) cases is noticeable. As $V$ continues to
increase, the  upper limit of the low energy branch, $-\Delta_-$, eventually 
reaches the value $\Delta_0$ (this occurs at $v=2$), and then NT from S to S'
becomes possible. Since the increase in current is mostly channeled through the
new NT channel, $dI/dV$ shifts to a `normal' value of $2e^2/h$.

For high values of $Z$, the regime with only AT at the NS interface is
bypassed, and the system jumps directly to GS. However, we have seen in the
previous paragraph that the new flow of quasielectrons can only be Andreev
reflected at SS'. The greater likelihood of AR causes a sharp increase in the
current, as shown in Fig. 5.

We find it remarkable that, as compared with the NS' (BTK) 
case, {\it the presence of a narrow S segment between the 
N lead and the wide S' reservoir enhances the current 
for moderate-to-large voltage values}. In particular, Fig. 5 shows that the
{\it excess  current} due to Andreev reflection can be {\it doubled} in the
case of transmissive  contacts. The experimental observation of this effect
would provide a proof of the existence of gapless
superconductivity in the clean superconductor S.

\section{I--V for NSN} 
 
We have seen that the self-consistent I--V curves for  a NS interface change
qualitatively if the narrow S wire is made finite and a wide superconducting
terminal is added  on the other side. We will see in this section that the
spectacular  behavior found in section V does not survive either in the other
natural realization of the NS interface, which consists in attaching a second
normal lead on the opposite side of the superconductor. For simplicity, we
consider a symmetric NSN structure, with the same value of $Z$ at the two NS
interfaces.  In such a case, the potential drops evenly at both interfaces and
the existing electron-hole symmetry (the flux of incoming electron from the
left mirrors that of holes from the right) simplifies the theoretical analysis.
The more complicated case of an asymmetric NSN structure is studied in the next
section.

The numerical results for a symmetric NSN are shown in Figs. 6 and 7. Non
self-consistent curves with analogous scattering assumptions are also shown.
For $Z=0$, the I--V curve is identical to that of a perfect normal wire,
namely, $I=(2e^2/h)V$. The reason is that a superconducting segment with fully
transmissive NS interfaces cannot improve on the conductance of a perfect N
wire,\cite{hui93} because it cannot make the average transmission $T_0 > 1$ (see
Eq. 5). Being $r=0$, AT is forbidden, and GS is achieved at $v = 2$. This is
clearly observed in the plot of $I_{\rm qp}$, 
which becomes positive. Remarkably,
the variation of the condensate current is such that the total current is
unaffected by the onset of GS, showing the behavior of a perfect N wire at all
voltages. Fig. 7 shows that $v_s$ and  $|\Delta|$ are also unaffected by the
presence of GS: $v_s$ displays the same linear increase with voltage and
$|\Delta|$ remains independent of it. The reason for the latter is that, as the
new unconventional branch emerges (see the inset of Fig. 2c), both its
quasilectron and quasihole states are occupied with very high probability (due
to NT at the interfaces), and this contributes to reinforce the order parameter
$|\Delta|$. However, this argument would merely explain an absence of strong
reduction. The reason why $|\Delta|$ remains exactly constant has to do with
the effective Galilean invariance of transport through a perfect wire. This is
obviously a nonequilibrium effect (see section IV) characteristic of structures
with high transmission.  Interestingly, the same prediction for the total
$I$-$V$ curve is obtained from a non self-consistent calculation with $v_s=0$. 
For $Z=0.5$, AT opens at $v \simeq 1.3$, as signaled by the current jump in
Fig. 6a and the negative values of Fig. 6b.  The enhancement of current is more
abrupt and occurs at lower voltages than in the non self-consistent
calculation. As for the NS structure, the negative values of $I_{\rm qp}$ are
more than compensated by the strong increase in $I_{\rm c}$ that is required to
accommodate the extra Cooper pairs transferred in AT events. A small reduction
of the gap (caused by the new presence of standard quasiparticles) is
compensated by a jump in $v_s$. GS is reached at $v \simeq 1.8$, but, as in the
transparent case, it is barely noticeable in the plot of the total current.
In the regime between the onsets of AT and
GS, possible impurities in the superconductor can only induce Andreev
reflection of the quasiparticles. This results in an insensitivity (within this
voltage range) of charge transport to the presence of impurities.
\cite{sanc96}

For higher values of the barrier strength ($Z=1$ and $Z=2$), the jumps in the
total current (at $v \simeq 1.7$ and 1.9, respectively) are much sharper than
in the low $Z$ cases. Comparison of Figs. 3b and 6b reveals that, as the
voltage increases, the AT regime is bypassed and the system jumps directly to
the GS state. A plausible reason for this behavior is the following: As the AT
mechanism opens at both interfaces, the two conventional branches become
strongly populated, which causes a reduction of the gap. On the other hand,
frequent AT events force the condensate to carry an intense positive current,
so that it can accommodate the extra flow of Cooper pairs. 
The two conditions (low
$|\Delta|$, high $I_{\rm c}$) can only be satisfied with a high value of $v_s$
that in turn gives entrance to GS, which is what we observe. In other words,
the AT regime is unstable for these structures, because it can only operate
with values of $v_s$ so high that the system shifts instead to GS. Fig. 7 shows
a jump in $v_s$ and a decrease in $|\Delta|$ that corroborate the picture
proposed above.

An interesting feature is that, at high voltages (GS state), $|\Delta|$ becomes
smaller for higher $Z$. The reason for this is that, as $Z$ increases, inner NR
at the interfaces becomes very important and an effective directional
randomization takes place in S, with a resulting equipartition of the four
quasiparticle channels. However, because {\it only one half of the incoming
scattering channels are occupied} (namely, electrons from the left and holes
from the right), unitarity requires that {\it equipartition in S can only be
achieved with} $\tilde{f}_{\lambda} \simeq \frac{1}{2}$
(see Eq. 17), which necessarily leads to a
strong reduction of the order parameter. In addition to this, in a realistic
physical scenario, an increase of quasiparticle confinement would be
accompanied by a stronger relaxation and an eventual thermalization. This
effect would contribute to suppress $|\Delta|$ even further. 

The dependence of the position of the first peak in the differential
conductance with temperature and barrier strength has been studied in Ref.
\cite{sanc95} There, it is shown that the FPDC shifts to lower voltages with
increasing temperature and decreasing $Z$. More structured
predictions about the
behavior of the FPDC are made later in section IX
for the case of an asymmetric NSN structure.

\section{NSN with many modes} 

So far we have employed a one-dimensional model for our calculations. We wish
to discuss the robustness of the main transport features against the existence
of many propagating modes at the Fermi energy. Specific numerical results are
presented in Fig. 8 for the case in which 20 transverse channels are available
for propagation.  We assume perfect interfaces so
that the transverse quantum number is conserved. 
The different modes are not, however, totally
independent, since  propagation through them is sensitive to the values of $q$
and $|\Delta|$, which depends in turn on the occupation of the whole set of
modes.

We observe that the current jumps that appeared in the one-channel case are
still present in a many mode context and occur at similar voltage values. The
physical reasons for the existence of these peaks are the same as in the
one-dimensional model. However,  the fact that the current enhancement takes
place within a very narrow voltage range is not obvious if one notes that,
nominally, the various thresholds should lie at different values of the voltage
for each of the 20 available modes, since each has a different longitudinal
$k_F$.  Within a hard wall model for the transverse potential, the quantities
$|\Delta| \pm \hbar v_F q$,   which determine the thresholds in the one-channel
case \cite{sanc95,mart95} must be replaced by $|\Delta| \pm  f(m) \hbar v_F q$,
where 
\beq 
f(m)=\left[1-\left( \frac{m}{w}\right) ^2\right] ^{1/2}, 
\eeq 
$w=2W/\lambda_F$ ($W$ is the device width), and $m$ runs from 1 to the total
number of  channels. As $V$ increases, quasiparticle transmission will occur
first for the mode with the highest longitudinal energy (here, that with
$m=1$), and one would expect naively that the other modes would open at
correspondingly higher voltage values. Inspection of Fig. 8 shows, however,
that well-defined jumps in the current take place within a  very narrow voltage
range, and in just one or a few steps. This points to the existence of a {\it
cascade effect} whereby, as one mode enters the new regime (AT or GS), it
induces a small sudden jump in $v_s$ (see Fig. 6a for the single channel case)
which helps anticipate the onset of the same regime for the following mode with
smaller $f(m)$, which in turn will favor the entrance of the following mode,
and so on.  A plot of $I_{\rm qp}$, $v_s$ and $|\Delta|$ (not shown) reveals
that, for these quantities, the transition takes place in a slightly wider
voltage range, and in a higher number of steps, although the naive expectation
of 20 different steps is never observed. This points to a preference of the
system for certain values of the total $dI/dV$ from which it departs in 
a very small voltage range. 

Another general feature is that current jumps occur at higher voltages in the
many-mode case. This is caused by the relative weight of channels with a
smaller longitudinal $k_F$ vector, which need higher $q$'s to undergo the
same transition. A complementary reason is that modes with smaller
$k_F$ are more strongly reflected at the interfaces and thus see a higher
effective $Z$, which also tends to raise the voltage thresholds. For similar
reasons, the presence of many modes tends to reduce  the amplitude of the order
parameter.

We conclude that the existence of a well-defined FPDC
is preserved in a many-channel context because of its fundamental
connection to the condensate flow. For this reason, 
the shift to lower voltages of the FPDC must be a clear
signature of  a nonzero superfluid velocity.

\section{Asymmetric NSN} 
 
In Fig. 9, we present curves for an asymmetric NSN junction where one of the 
barriers is fixed at a low, nonzero value of  $Z$ (namely, $Z_2$=0.5), while
the strength $Z_1$ of the other barrier varies from small to large values. 
The upper part of Fig. 9 represents  the differential conductance for
both the self-consistent and the non self-consistent calculations, while the
lower part shows the relevant energies of the   problem, namely, the voltage
differences at each interface, the  magnitude $|\Delta|$ of the self-consistent
order parameter, and  the thresholds $\Delta_{\pm}$. The calculations  have
been performed for a finite temperature of $T=2$ K, with values of $E_F$,
$\hbar \omega_D$, and $\Delta_0$ which correspond to Pb, whose critical
temperature is $T_c=7.2$ K. Because the voltage does not drop symmetrically at
the two interfaces, the intermediate voltage at the superconducting segment is
a third parameter (besides $q$ and $|\Delta|$) that must be determined
self-consistently. A practical method consists in  requiring the current to
be the same in the two N leads and adjust the voltage in S (for given $q$ and
$|\Delta|$) until this is achieved. With fixed voltage drops, the values of $q$
and $|\Delta|$ are in turn determined  as in the previous sections. We have
found that, usually, $I_N$ does not vary much in this second adjustment.

The case $Z_1 \gg Z_2$ (Figs. 9d and 9e) is easy to understand. Because of its
poor transmittivity, interface 1 acts as a bottle neck for current flow, and
its properties, which are those of a NS tunnel junction, determine the global
transport behavior. Because transmission is low, the inclusion of
self-consistency does not introduce changes in the position of the first peak,
although it affects its height. As in BTK, the first peak lies,
characteristically, at $V \simeq \Delta_0/e$ in both cases. Fig. 9c shows that,
when the two barriers are identical, one obtains the largest differences
between the two descriptions, with the peaks located in quite different
positions. By symmetry, the non self-consistent peak must be at $V \simeq
2\Delta_0/e$ (i.e., when the voltage drops by $\Delta_0/e$ at each interface).
The inclusion of a finite condensate flow gives rise to the lowering of the
voltage threshold and, for this reason, the FPDC tends to stay in the
neighborhood  of $V \simeq \Delta _0/e$. Finally, when $Z_1 \ll Z_2$ (Figs. 9a
and 9b),  the bottle neck shifts to the second interface. However, since this
is already quite transmissive ($Z_2=0.5$), we do not reproduce the transport
behavior of a NS tunnel junction. Again, the presence of $v_s$ is responsible
for important differences. While the BTK-type calculation tends to send the
peak back to the neighborhood of $V \simeq \Delta_0/e$ (as $Z_1$ decreases), 
the self-consistent calculation predicts a peak position below this threshold
(see also the curve for $Z=0.5$ in Fig. 3). The evolution of the FPDC as a
function of $Z_1$ is summarised in Fig. 10.

The lower curves of Fig. 9 show a correlation between the apparition of the
first peak in the self-consistent $dI/dV$ and the reaching of the $\Delta_-$
threshold by the voltage drop at the interfaces. As AT opens,
there is a corresponding enhancement of the current.

The continuous evolution from small to high $Z_1$ could be realized
experimentally by means of a STM measurement. A normal tip could be applied to
a superconducting particle located on top of a normal substrate.  $Z_1$ could
be varied by changing the tip-particle distance, and the $Z_2$ value
characterizing the island-substrate contact would be a fixed parameter. The
same experiment could be performed on S islands of different sizes  with
different levels of current dilution. One expects transport through broad
islands to be well described by a non self-consistent calculation, while the
effect of a condensate flow should be essential to understand the transport
behavior of a narrow island.

\section{Conclusions}

We have computed the self-consistent current-voltage characteristics of several
types of hybrid normal-superconductor structures. We have focussed on the new
physical features arising from the implementation of  self-consistency and from
the related condition of current conservation ($I_N=I_S$). We have shown on
rather general grounds that the self-consistency effects are important in
transmissive structures where current is not diluted in the superconductor.

The existence of current flow in the superconducting condensate ($v_s \neq 0$)
introduces qualitative changes in the scattering of quasiparticles, affecting
even the energy thresholds for propagation. The new structure of scattering
channels gives rise to a rich set of transport regimes, including voltage
ranges in which only Andreev transmitted quasiparticles can enter the
superconductor or in which a peculiar form of gapless superconductivity exists.
We have seen that, when applied to the isolated NS interface, the
implementation of self-consistency predicts remarkable transport properties
which, however, are not observed in practice if proper boundary conditions are
assigned to the superconductor side. We have explicitly shown that, if another
normal lead, or a wide superconductor (for which $v_s\simeq 0$), is attached on
the opposite side of S, then the predicted I--V curves vary qualitatively. The
two main features are (i) a lowering of the voltage at which the first peak in
the differential conductance occurs, and (ii) a global enhancement of current
in a wide range of voltages due to the  increased availability (favored by
$v_s\neq 0$) of charge-transmitting channels. In the case of NSN, we have shown
these effects persist in the presence of many propagating channels at the Fermi
energy. Predictions (i) and (ii) are quite robust and should be observable. The
effect of current increase should be particularly clear in NSS' structures,
where S is narrow and S' is wide, in the form of an  enhancement of the excess
current. Such an effect would be a strong indication of the existence of gapless
superconductivity in the narrow superconductor. For asymmetric NSN structures
in which at least one of the barrier strengths can be tuned at will, we suggest
experiments which would permit the identification of finite condensate flow
effects.

The implementation of self-consistency in transport calculations puts a strain
on the numerical resources and necessarily reduces the range of situations one
can explore. By introducing some approximations (see the Introduction), we have
reduced the extra numerical requirements to a minimum. The simplicity
of the resulting model has allowed us to identify the essential
physical features associated to current conservation. A shortcoming is that the
predictions obtained, although rich in structure, are of semiquantitative
character. The development of numerical methods to compute transport properties
at finite current densities
with quantitative predictive power for interesting structures is an important
challenge. Together with the work of Refs. \cite{sanc95,mart95}, we have
attempted here to give some steps in this direction. We hope that the present
work will stimulate more research on the rich physics of current-conserving
transport in normal-superconductor structures.

\acknowledgments 
 
We wish to thank C.W.J. Beenakker, F. Beltram, J. Ferrer,  C.J. Lambert,  A.
Martin,  J.G. Rodrigo, M. Poza,  and G. Sch\"on  for valuable discussions. 
This project  has been supported by Direcci\'on General  de Investigaci\'on
Cient\'{\i}fica  y T\'ecnica, Project no. PB93-1248, and by  the HCM Programme
of the EU. One of us (J.S.C.) acknowledges the support from Ministerio de 
Educaci\'on y Ciencia through a FPI fellowship.\\ 

\appendix 
\section{ASYMPTOTIC SELF-CONSISTENCY} 
 
A quasiparticle of energy $E$ approaching the structure through channel
$\alpha$ can be scattered into any of the outgoing channels $\beta$ with
probability amplitude $S_{\beta \alpha}$. Far enough from the scattering region
($|x| \rightarrow \infty$), the wave function is of the form
\beqa 
\psi_{\alpha}(x)= 
\left[ \begin{array}{c} u_{\alpha}(x) \\ 
v_{\alpha}(x) 
\end{array} \right]=\phi_{\alpha}(x)+\sum_{\beta}
\left(\frac{\nu _{\alpha}}{\nu _{\beta}} \right)^{1/2} 
S_{\beta \alpha} \phi_{\beta}(x), 
\eeqa 
where $\nu _{\lambda}$ is the group velocity of  channel $\lambda$. The
scattering channels are the free propagating solutions for a perfect
superconductor with Cooper pair momentum $2q$. The wave function for channel
$\lambda$ is \cite{dege66}
\beqa 
\phi_{\lambda}(x)= 
\left[ \begin{array}{c} u_{\lambda}e^{iqx} \\ 
v_{\lambda}e^{-iqx} 
\end{array} \right]e^{ik_{\lambda}x} \eta_{\lambda}(x), 
\eeqa 
where  $\eta_{\lambda}(x)$ is an indicator function taking value 1 if point $x$
lies in the lead of channel $\lambda$ and 0 otherwise.
$u_{\lambda}$, $v_{\lambda}$ are the coherence factors for vector $k_{\lambda}$
(see Eq. 11).  For greater clarity, we refer to vectors of incoming channels
$\alpha$ with letter $k$ and those of outcoming channels $\beta$ with $p$.
Inserting Eqs. (A1-A2) into Eq. (2) with $\Delta(x)=|\Delta|e^{2iqx}$, we
obtain
\beqa 
|\Delta| &=& \frac{g}{h}\int_0 ^{\hbar \omega _D} dE
\sum_{\alpha} \frac{(1-2f_{\alpha})}{\nu _{\alpha}}
\left[ v_{\alpha} 
e^{-ikx}\eta_{\alpha}(x)+\sum_{\beta} 
\left(\frac{\nu _{\alpha}}{\nu _{\beta}} \right)^{1/2} 
S^* _{\beta \alpha}v_{\beta} 
e^{-ipx}\eta_{\beta}(x) \right] \times \nonumber \\ 
&\times& \left[ u_{\alpha}e^{ikx}\eta_{\alpha}(x)+ 
\sum_{\beta '}
\left(\frac{\nu _{\alpha}}{\nu _{\beta '}} \right)^{1/2}
S_{\beta ' \alpha}u_{\beta '} 
e^{ip 'x}\eta_{\beta '}(x) \right]. 
\eeqa
To achieve asymptotic self-consistency, we take the limit
$|x|\rightarrow\infty$ and neglect two types of terms: (i) $e^{i(p ' - p)x}$
with $\beta \neq \beta '$, 
which goes roughly as $e^{2ik_Fx}$ and yields unphysical 
oscillations in $|\Delta|$ at the scale of the  Fermi wavelength; (ii)
$e^{i(k-p)x}$ appears because of the electron-hole coherence between $\alpha$
and $\beta$. Since  $k$ and $p$ take values within an  interval of width
$\xi_0^{-1}$, the integrated term becomes negligible for $|x| \gg \xi_0$
(energy is integrated within and effective 
interval of order $\Delta_0$). Thus, we may
write
\beqa 
|\Delta|=\frac{g}{h}\int_0^{\hbar \omega_D}dE 
\sum _{\alpha}(1-2f_{\alpha}) \left [ 
\frac{v_{\alpha}u_{\alpha}}{\nu_{\alpha}} 
\eta_{\alpha}(x)+\sum_{\beta} 
\frac{v_{\beta}u_{\beta}}{\nu _{\beta}} 
|S_{\beta \alpha}|^2 \eta_{\beta}(x) \right ]. 
\eeqa
Noting that $\tilde{f}_{\beta}=\sum_{\alpha}f_{\alpha}  |S_{\beta \alpha}|^2$
and $\tilde{f}_{\alpha}=f_{\alpha}$, as well as the unitarity condition
$\sum_{\alpha}|S_{\beta \alpha}|^2=1$, we are able to reproduce Eq. (12) of the
main text. Eqs. (13-14)  can be easily obtained with the same set of
assumptions.

\begin{figure}[h]
\centerline{
\psfig{figure=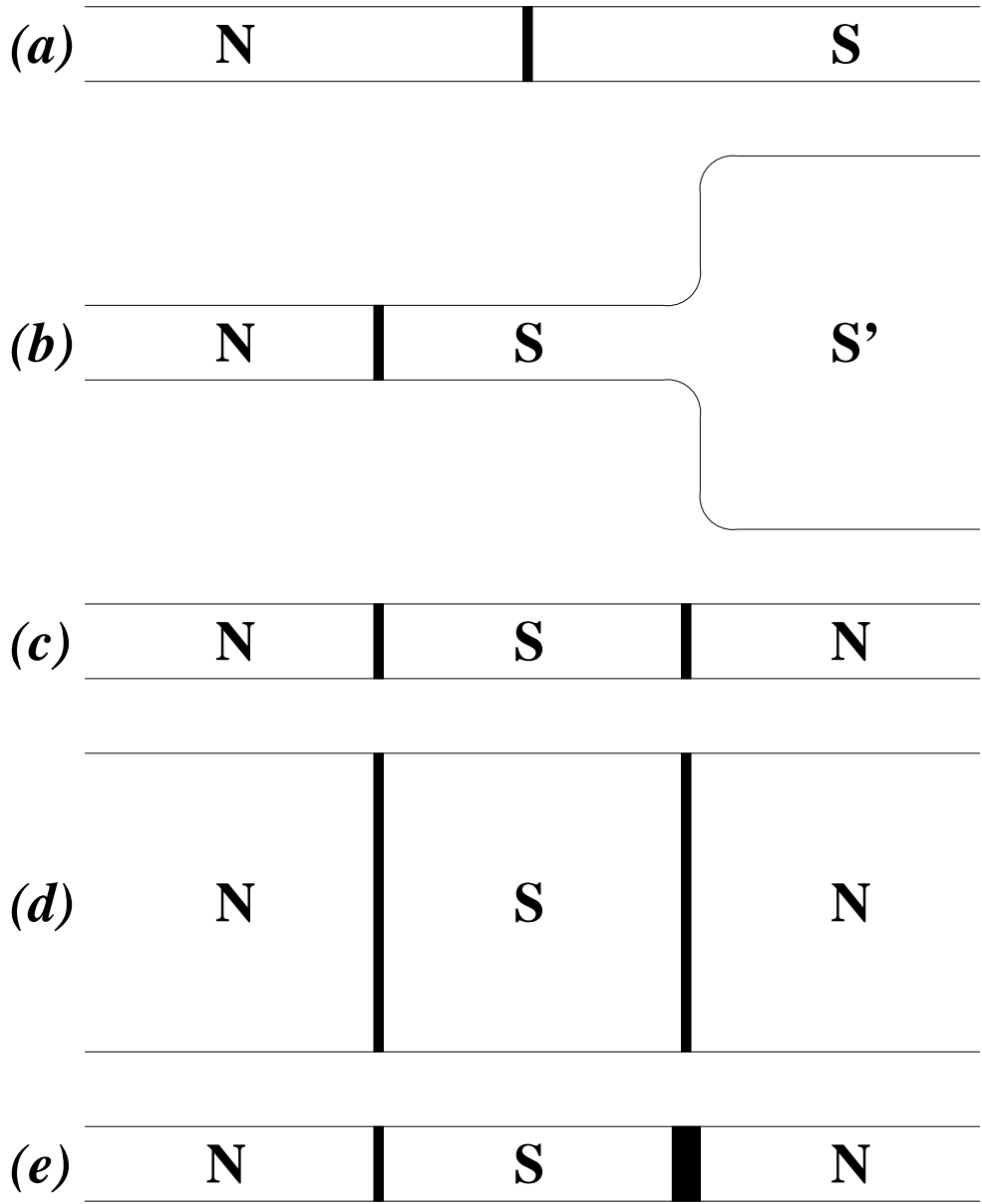,height=16cm}
}
\vspace{2cm} 
\caption{ 
Schematic representation of the structures to be studied:
(a) Normal-superconductor junction (NS), 
(b) NSS' structure with S' a wide superconductor for which $v_s \simeq 0$,
(c) symmetric NSN structure,  
(d) symmetric NSN with many channels, and 
(e) asymmetric NSN (with a different barrier strength at each 
interface). 
} 
\end{figure}
 
\begin{figure}[h]
\centerline{
\psfig{figure=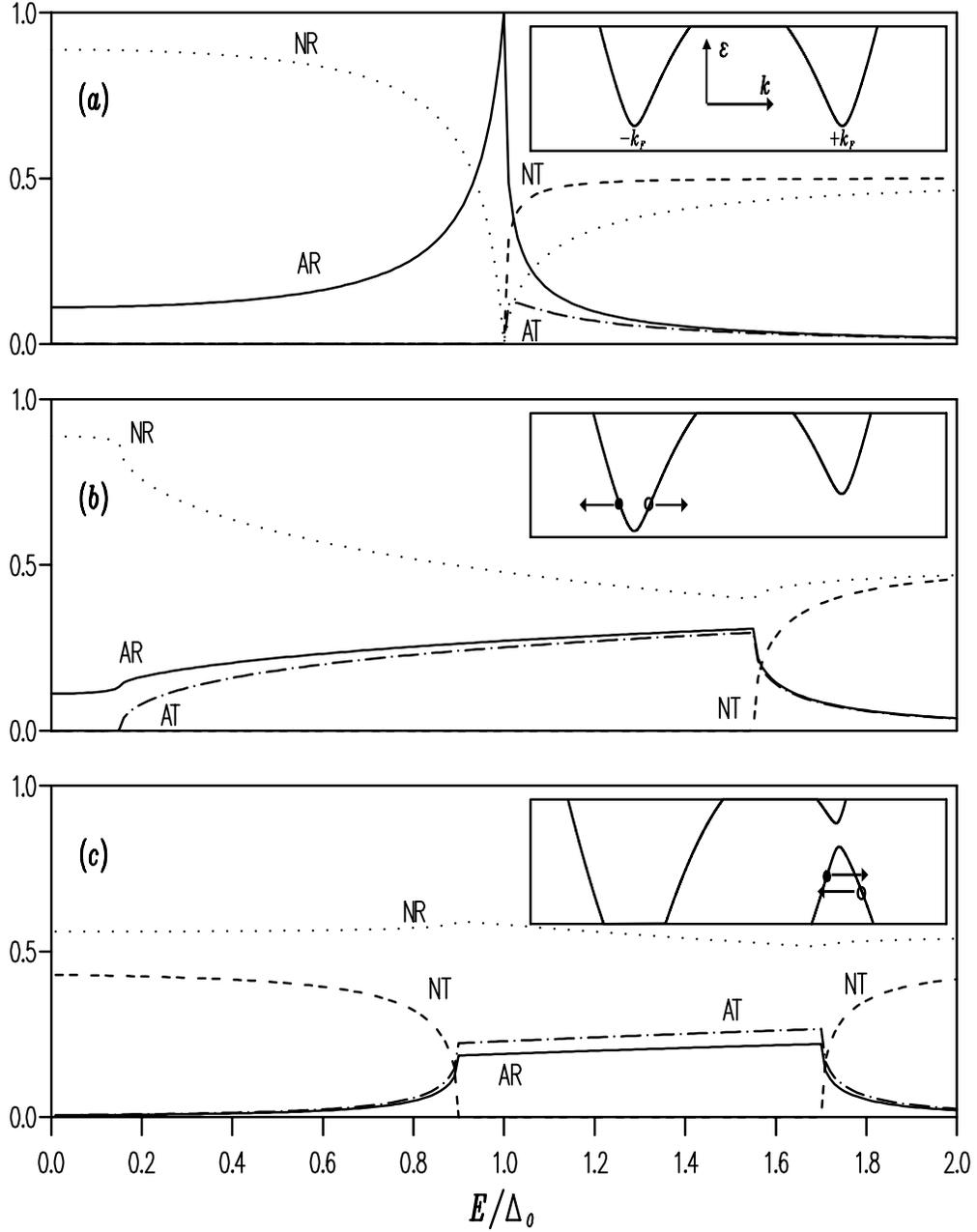,height=20cm,angle=180}
}  
\caption{ 
Scattering probabilities for an electron incident on
a NS junction from the normal lead. The superconducting 
condensate has a finite $v_s$ and $Z=1$ is taken for 
the barrier at the interface.
(a) $v_s/v_d=0$ (BTK);  
(b) $v_s/v_d=0.7$; and 
(c) $v_s/v_d=1.3$, GS regime, in which normal 
transmission at low energies is possible (here,
$Z=1.1$ has been taken to show AR and AT probabilities
separately). Insets: Schematic 
quasiparticle dispersion relation, $\varepsilon(k)$,  
for each case. Filled 
(empty) circles indicate electron- (hole-) like propagation. 
} 
\end{figure} 
 
\begin{figure}[h]
\centerline{
\psfig{figure=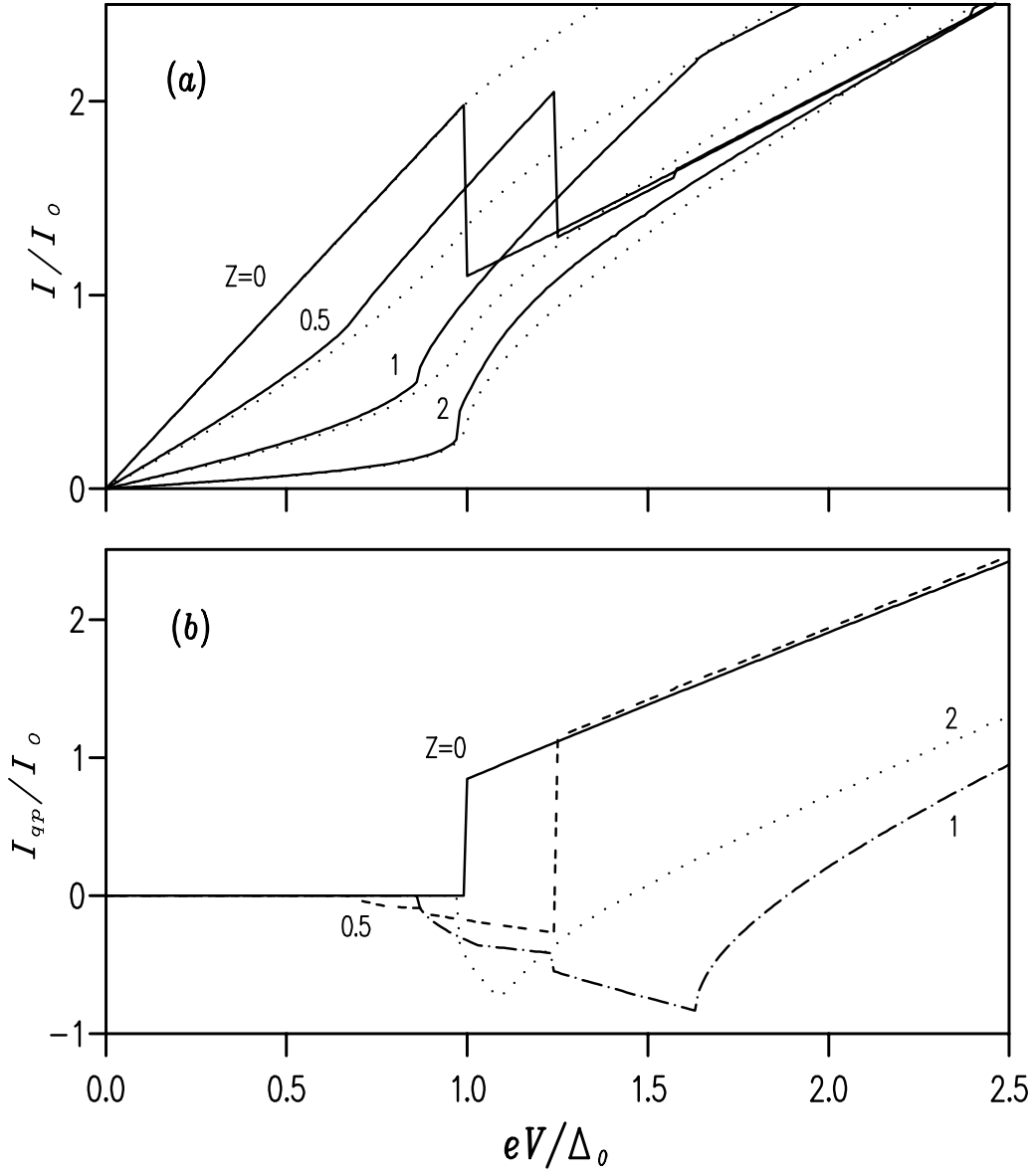,height=20cm,angle=180}
}  
\caption{ 
I--V characteristics of a NS junction for 
different values of the effective barrier 
strength ($Z=0$, $0.5$, $1$ and $2$).
The current is given in units of
$I_0=2e\Delta _0/(h(1+Z^2))$ 
(a) Total electrical current; 
solid (dotted) lines are obtained from a self-consistent 
(non self-consistent) calculation. 
(b) Quasiparticle component of the self-consistent  
current. 
} 
\end{figure} 
 
\begin{figure}[h]
\centerline{
\psfig{figure=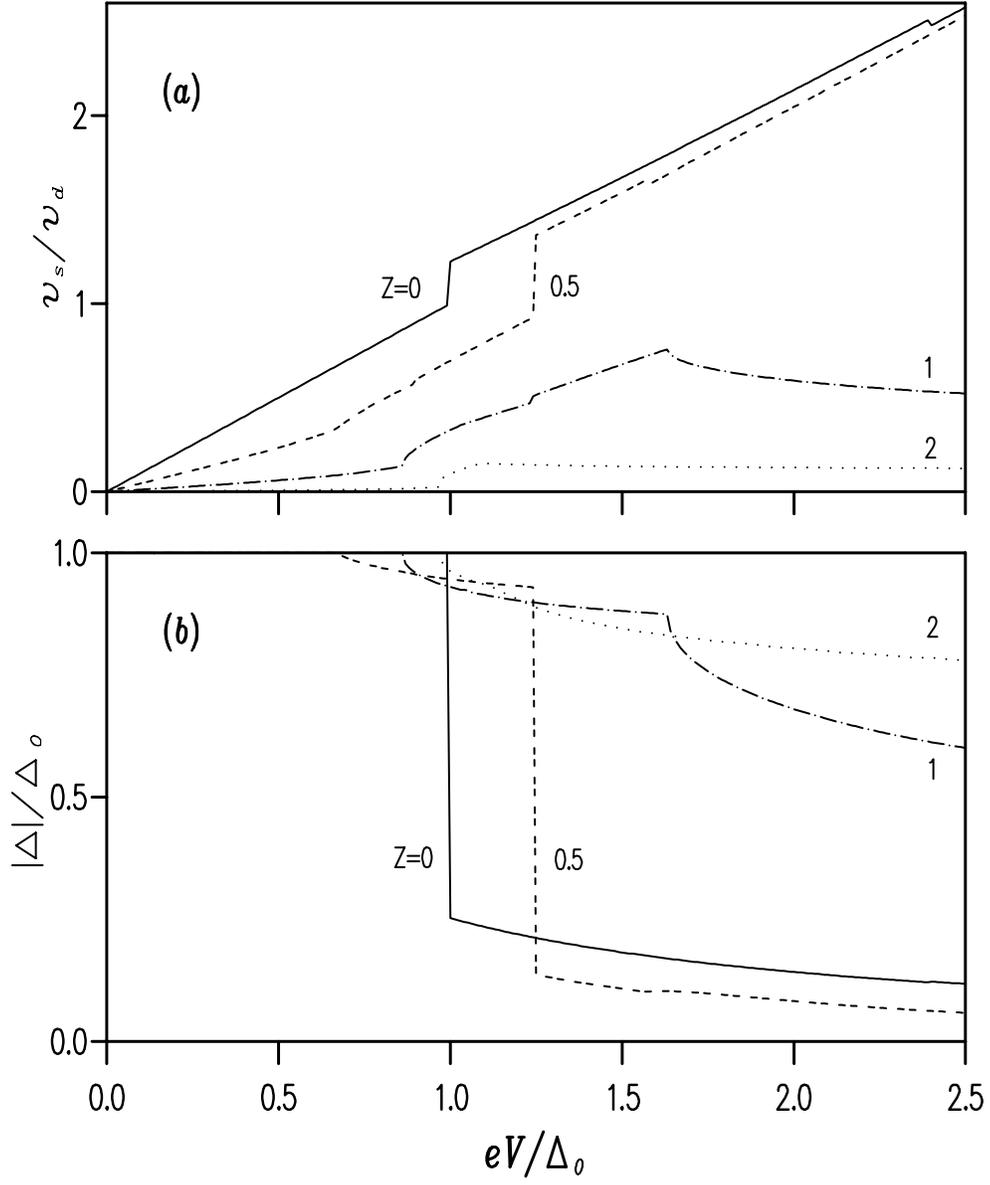,height=20cm,angle=180}
} 
\caption{ 
Self-consistent parameters for the curves 
of Fig. 3:
(a) superfluid velocity in units of the depairing 
velocity $v_d$; 
(b) magnitude of the order parameter in units of 
$\Delta_0$ 
} 
\end{figure} 
 
\begin{figure}[h]
\centerline{
\psfig{figure=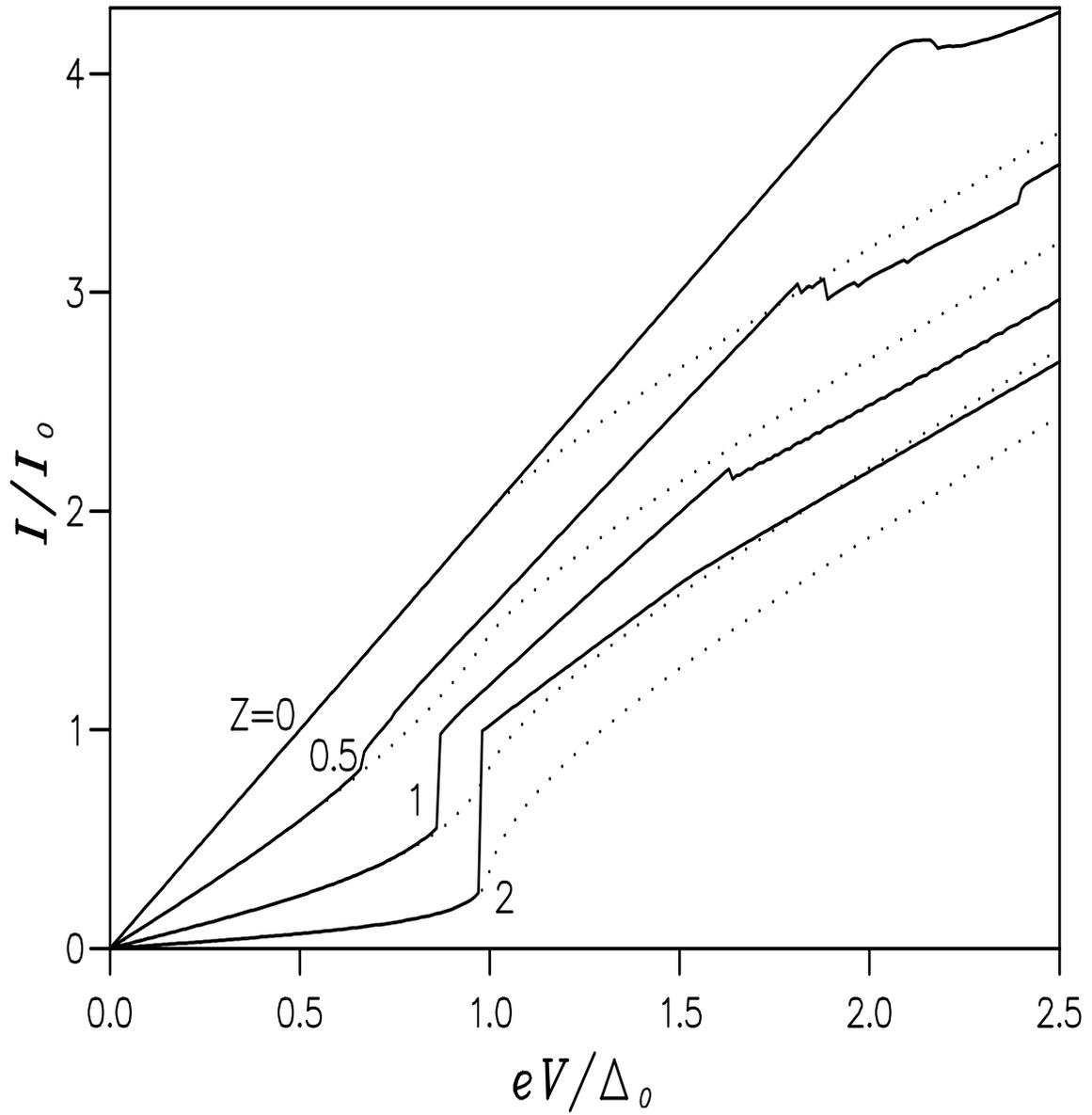,height=20cm,width=18cm,angle=90}
} 
\caption{ 
I--V characteristics of a NSS' structure 
for $Z=0$, $0.5$, $1$ and $2$ at the NS interface 
(solid lines). Dotted lines correspond to the NS'  
(non self-consistent NS) case. Incoherent scattering has been 
assumed. 
} 
\end{figure} 
 
\begin{figure}[h]
\centerline{
\psfig{figure=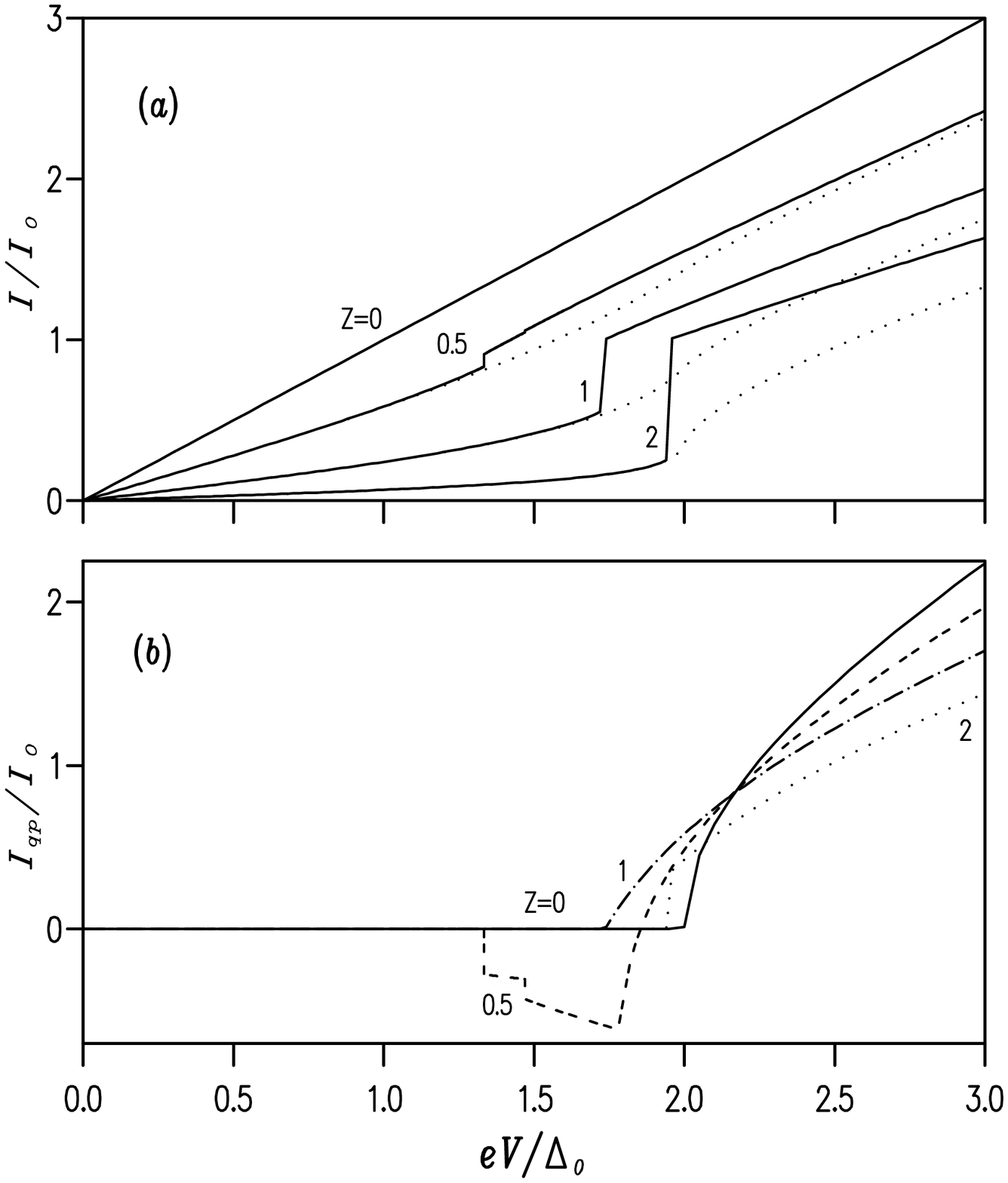,height=20cm,angle=180}
} 
\caption{ 
Same as Fig. 3 for a symmetric NSN structure.
} 
\end{figure} 
 
\begin{figure}[h]
\centerline{
\psfig{figure=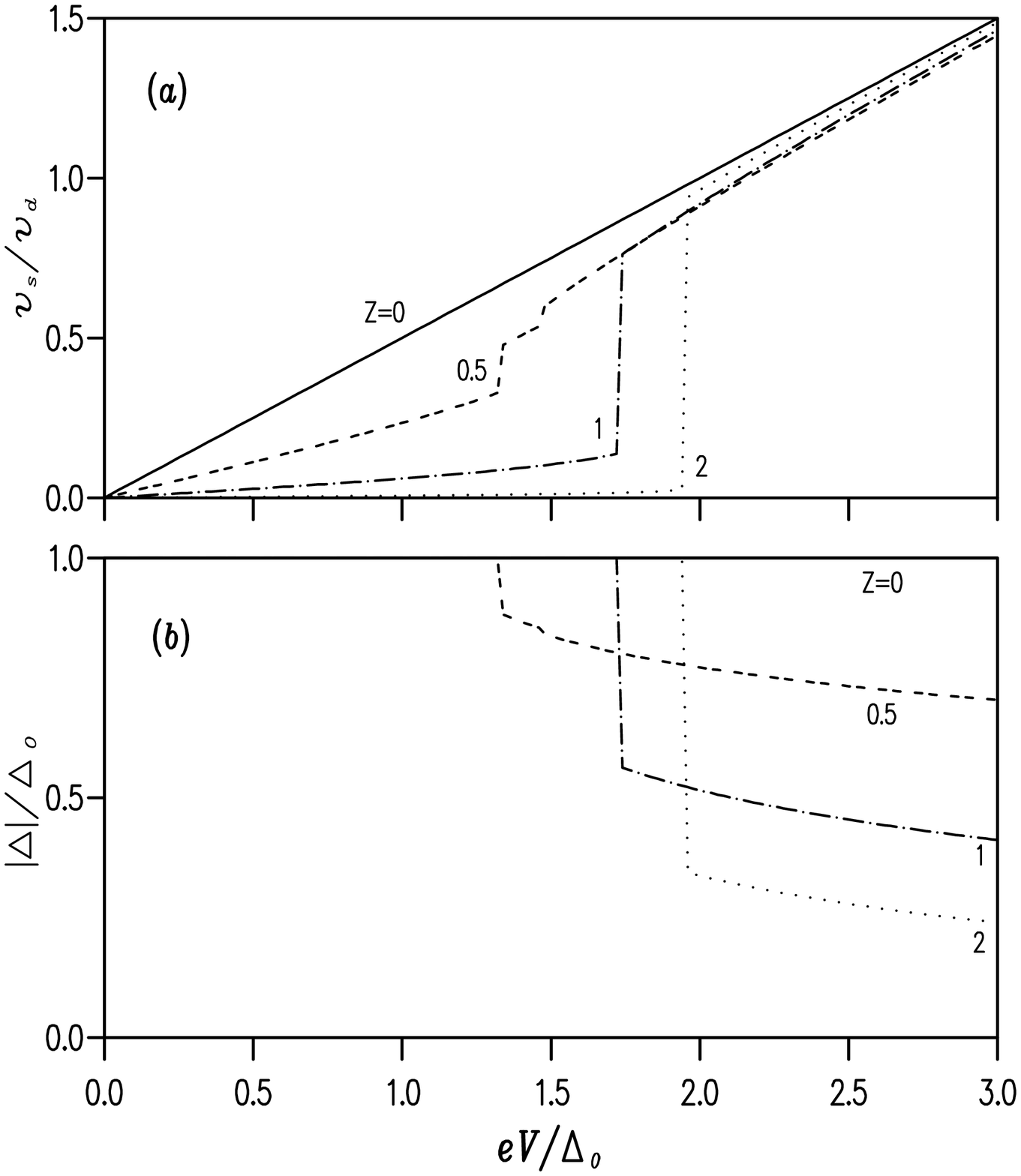,height=20cm,angle=180}
} 
\caption{ 
Same as Fig. 4 for a symmetric NSN structure.
} 
\end{figure} 
 
\begin{figure}[h]
\centerline{
\psfig{figure=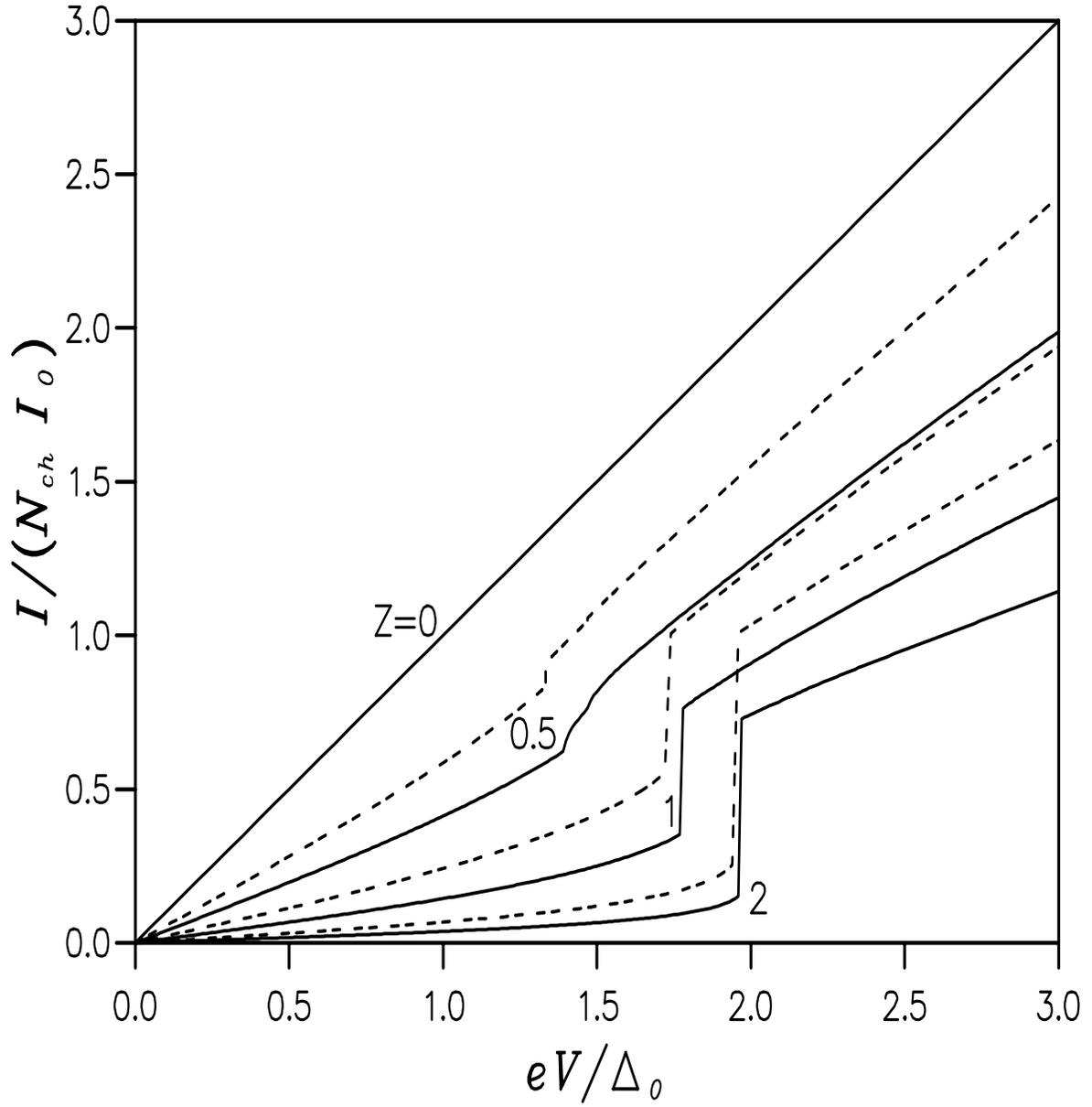,height=20cm,width=18cm,angle=90}
}  
\caption{ 
I--V characteristics for a symmetric NSN 
structure with 20 transverse mode (solid), 
convenient rescaled for comparison with the single channel case (dashed).} 
\end{figure} 
 
\begin{figure}[h]
\centerline{
\psfig{figure=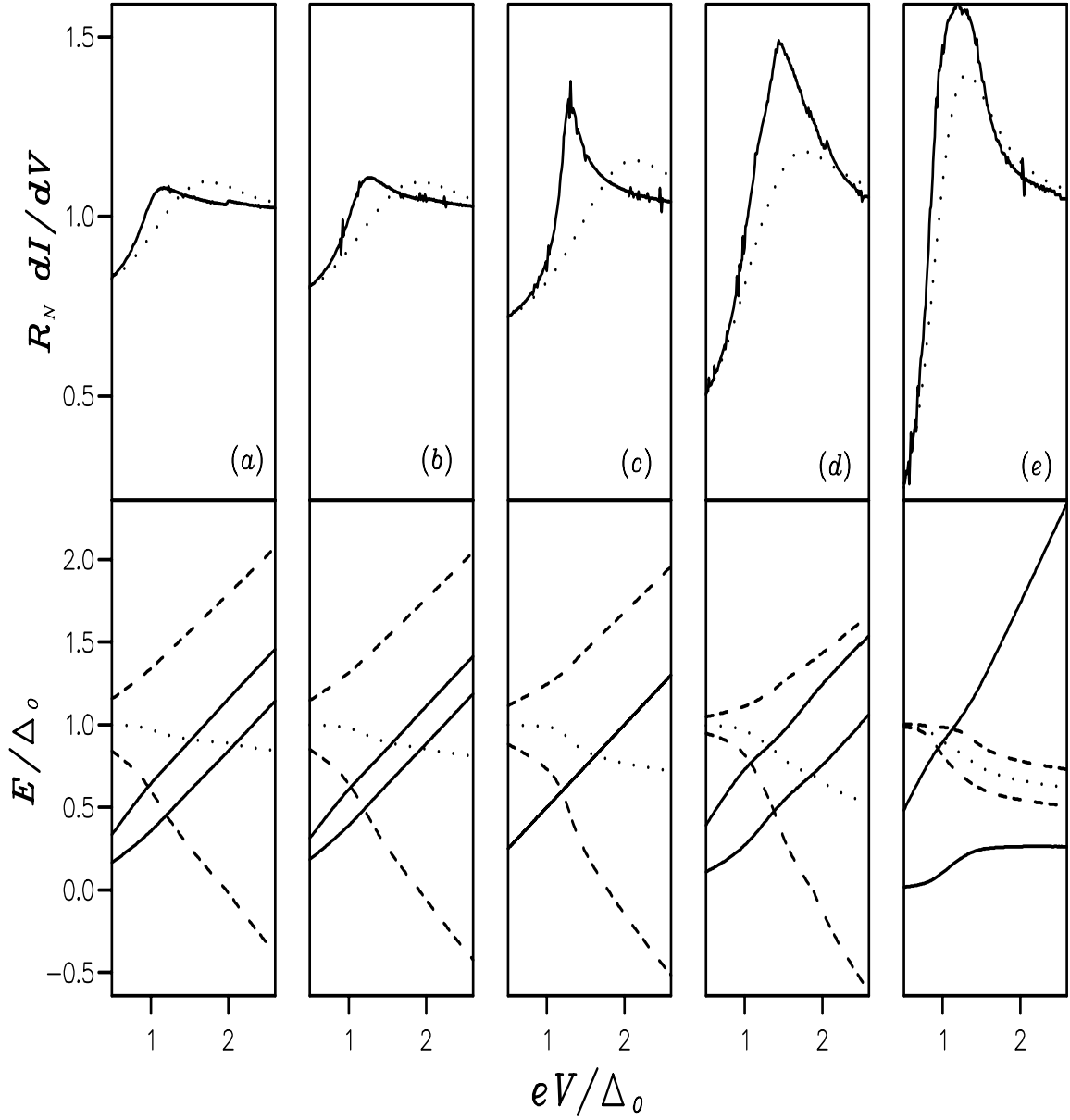,height=20cm,width=18cm,angle=90}
}  
\caption{ 
Asymmetric NSN structure with fixed $Z_2=0.5$
For S, data of Pb have been taken.
Temperature is $T=2$K ($\ll T_c$).
The upper part shows 
differential conductances for self-consistent 
(solid) and non self-consistent (dotted) 
calculations. The normal resistance is $R_N=(h/2e^2)(1+Z_1^2+Z_2^2)$. 
The lower part shows (in units of $\Delta_0$):
Voltage drops
at each interface (solid), 
magnitude of the order parameter $|\Delta|$ (dotted), 
and of
$\Delta_+$ and $\Delta_-$ thresholds (dashed). 
Values for $Z_1$ are: (a) 0.1, (b) 0.25, 
(c) 0.5, (d) 1, and (e) 2. 
} 
\end{figure} 
 
\begin{figure}[h]
\centerline{
\psfig{figure=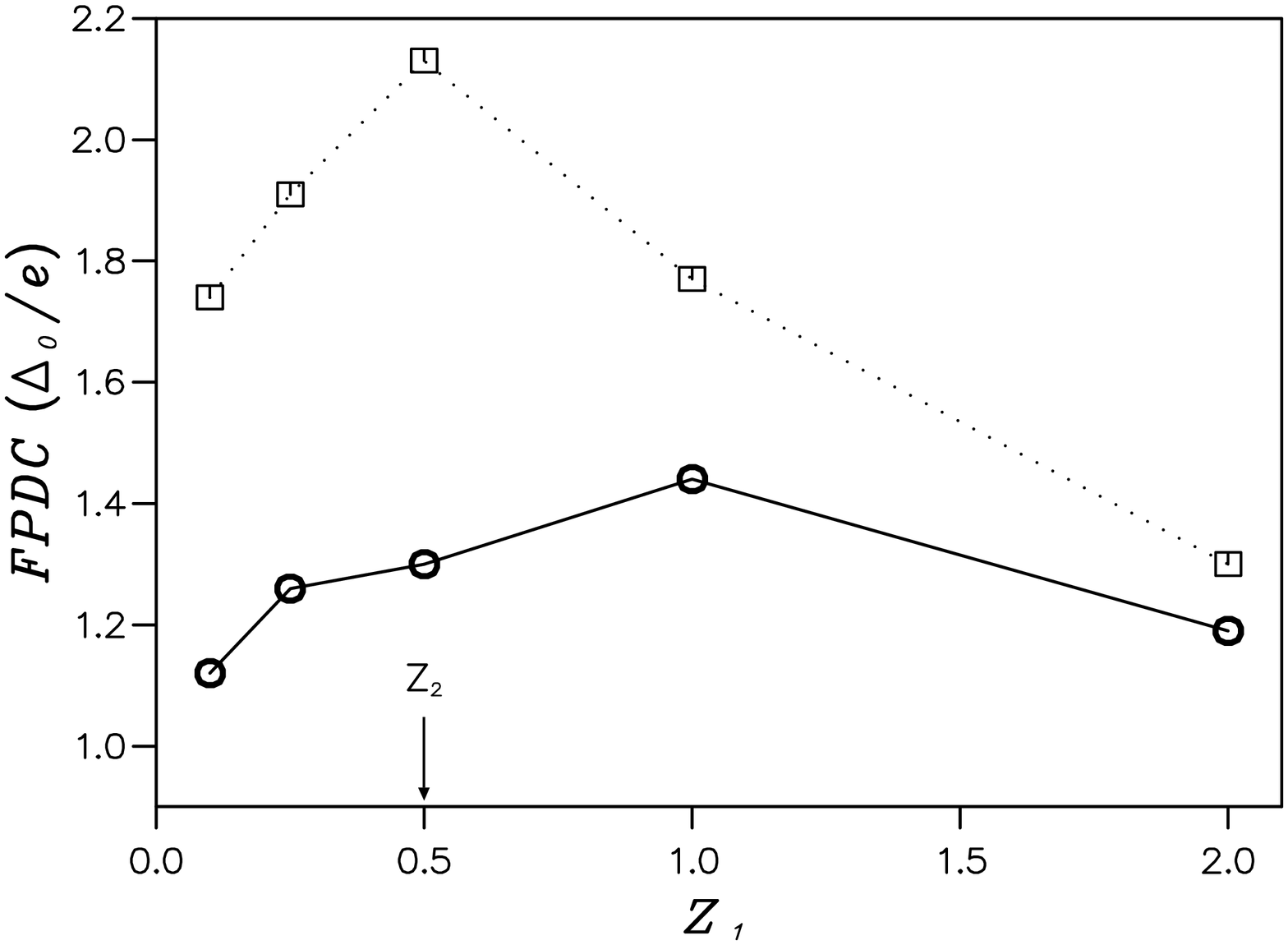,height=20cm,width=18cm,angle=90}
} 
\caption{ 
Evolution of the position of the first peak in the
differential conductance as a function of the barrier
$Z_1$ for a fixed $Z_2=0.5$. Self-consistent (non self-consistent)
results are represented with circles (squares).}
 
\end{figure}


\begin{references} 
 
\bibitem{hekk95} For an updated overview, see 
{\it Mesoscopic Superconductivity}, Proc. 
of the NATO-ARW, F.W.J Hekking,  
G. Sch\"{o}n, D.V. Averin, eds. (North-Holland, 
Amsterdam, 1995). 
 
\bibitem{andr64} A.F. Andreev, Sov. Phys. JETP {\bf 19}, 1228 
(1964).  
 
\bibitem{been95} See, for example, C.W.J. Beenakker, in {\it 
Mesoscopic Quantum Physics}, E. Akkermans, G. Montamboux, 
J.-L. Pichard, and J. Zinn-Justin, eds. (North-Holland, Amsterdam, 1995). 
 
\bibitem{dege66} P.G. de Gennes, {\it Superconductivity of Metals and 
Alloys} (Benjamin, New York, 1966). 
 
\bibitem{furu91} A. Furusaki and M. Tsukada, 
Solid State Commun. {\bf 78}, 299 (1991). 
 
\bibitem{bagw94} P.F. Bagwell, Phys. Rev. {\bf B49}, 6841 (1994). 
 
\bibitem{sols94} F. Sols and J. Ferrer,  
Phys. Rev. {\bf B49}, 15913 (1994).

\bibitem{ferr90} J. Ferrer, Ph.D. Thesis, Universidad Aut\'onoma de  
Madrid (1990), unpublished. 
 
\bibitem{baym61} G. Baym and L.P. Kadanoff, Phys. Rev. {\bf 124}, 
287 (1961); {\it ibid.} {\bf 127}, 1391 (1962). 

\bibitem{sanc95} J. S\'anchez-Ca\~nizares and F. Sols, 
J. Phys.: Condens. Matter {\bf 7}, L317 (1995). 
 
\bibitem{bard62} J. Bardeen, Rev. Mod. Phys. {\bf 34}, 667 (1962). 
 
\bibitem{tink96} M. Tinkham, {\it Introduction to superconductivity} 
(McGraw-Hill, New York, 1996).
 
\bibitem{blon82} G.E. Blonder, M. Tinkham, T.M. Klapwijk, 
Phys. Rev. {\bf B25}, 4515 (1982). 
 
\bibitem{klap94} A review of recent work on 
NS structures where N is a doped semiconductor 
has been given by T.M. Klapwijk, Physica B {\bf  
197}, 481 (1994). 
 
\bibitem{gray81} {\it Nonequilibrium 
Superconductivity, Phonons, and Kapitza Boundaries}, 
K.E. Gray, ed. (Plenum, New York, 1981), and 
{\it Nonequilibrium Properties of Superconductors  
(Transport Equation Approach)}, A.G. Aronov, Yu.M. 
Galperin, V.L. Gurevich, and V.I. Kozub, eds. 
(Elsevier, Amsterdam, 1986).

\bibitem{lamb91} C.J. Lambert, J. Phys.: Condens. Matter {\bf 3},
6579 (1991); {\bf 5} 707 (1993).

\bibitem{been92} C.W.J. Beenakker, Phys. Rev. B {\bf 46}, 12841 (1992).

\bibitem{taka92} Y. Takane and H. Ebisawa, J. Phys. Soc. Japan {\bf 61},
1685 (1992); {\bf 61}, 2858 (1992).

\bibitem{land57} R. Landauer, IBM J. Res. Develop. {\bf 1}, 223 (1957).

\bibitem{butt86} M. B\"uttiker, Phys. Rev. Lett. {\bf 57}, 1761 (1986).

\bibitem{mart95} A. Martin and C.J. Lambert, Phys. Rev. 
{\bf B51}, 17999 (1995).

\bibitem{mcmi68} W.L. McMillan, Phys. Rev. {\bf 175}, 559 (1968).

\bibitem{brud90} C. Bruder, Phys. Rev. B {\bf 41}, 4017 (1990).

\bibitem{sanc97} J. S\'anchez-Ca\~nizares and F. Sols, to be published.
 
\bibitem{been91} C.W.J. Beenakker and H. van Houten,  Solid State  
Physics {\bf 44}, 1 (1991). 
 
\bibitem{mels94} J.A. Melsen and C.W.J. Beenakker, 
Physica B {\bf 203},  
219 (1994).  
 
\bibitem{clar72} J. Clarke, Phys. Rev. Lett. {\bf  28}, 1363 
(1972); M. Tinkham and J. Clarke, {\it ibid.} {\bf 28}, 1366 (1972).

\bibitem{blam90} M.G.Blamire, E.C.G. Kirk, J.E. Evetts, and
T.M. Klapwijk, Phys. Rev. Lett. {\bf 66}, 220 (1991).

\bibitem{klap93} D.R. Heslinga and T.M. Klapwijk, Phys. Rev. B {\bf 47},
5157 (1993).
 
\bibitem{sanc96} J. S\'anchez-Ca\~nizares and F. Sols, 
J. Phys.: Condens. Matter {\bf 8}, L207 (1996).

\bibitem{peth79} C.J. Pethick and H. Smith, Ann. Phys. (N.Y.) {\bf 119},
133 (1979).
 
\bibitem{likh79} K.K. Likharev, Rev. Mod. Phys. {\bf 51}, 101 (1979).

\bibitem{lang67} J.S. Langer and V. Ambegaokar, 
Phys. Rev. {\bf 164}, 498 (1967). 

\bibitem{commscatt} Explicit expressions in terms of $Z$ for the particular 
symmetric case of a delta barrier have been given in Ref. \cite{blon82}.

\bibitem{commconv}
We follow the standard convention in which $\varepsilon > 0$. 
For each quasiparticle $n$ there is another solution $n'$ with energy
$\varepsilon_{n'}=-\varepsilon_n<0$ and wavefunction $(u_{n'},v_{n'})
=(-v_n^*, u_n^*)$.\cite{dege66} These two solutions are not independent, 
since creating quasiparticle $n,\sigma$ ($\sigma$ is the spin) is equivalent 
to destroying quasiparticle $n',-\sigma$.\cite{sanc95} More specifically,
$ \gamma^{\dagger}_{n\downarrow} =\gamma_{n'\uparrow}$, and
$\gamma_{n\uparrow}= - \gamma^{\dagger}_{n'\downarrow}$.

\bibitem{commgap} One may also state that, because it is the product
$u_{\lambda}v_{\lambda}(1-2\tilde{f}_{\lambda})$ that appears in the gap
equation,  the effect of occupied `unconventional' states is equivalent to that
of empty  `conventional' states, which is to strengthen the gap.

\bibitem{rodr94} J.G. Rodrigo, N. Agra\"{\i}t, C. Sirvent, 
and S. Vieira, 
Phys. Rev. {\bf B50}, 12788 (1994). 
 
\bibitem{poza95}  M. Poza, J.G. Rodrigo, and S. Vieira, 
Physica B {\bf 218}, 265 (1996). 
                        
\bibitem{hui93} V. Hui and C.J. Lambert, 
J. Phys.: Condens. Matter {\bf 5} L651 (1993). 
 
\end{references}
\end{document}